\documentclass[aps,twocolumn]{revtex4}
\usepackage{graphicx}
\usepackage{amsmath}

\usepackage{color}
\begin{document}
\title{Plasmon-enhanced spin-orbit interaction of light in graphene}
\author{A. Ciattoni$^1$}
\email{alessandro.ciattoni@aquila.infn.it}
\author{C. Rizza$^1$}
\author{H. W. H. Lee$^2$}
\author{C. Conti$^{3,4}$}
\author{A. Marini$^5$}
\email{andrea.marini@aquila.infn.it}
\affiliation{$^1$Consiglio Nazionale delle Ricerche, CNR-SPIN, Via Vetoio 10, 67100 L'Aquila, Italy}
\affiliation{$^2$Department of Physics, Baylor University, Waco, TX, 76798, United States}
\affiliation{$^3$Institute for Complex Systems (ISC-CNR), Via dei Taurini 19, 00185, Rome, Italy}
\affiliation{$^4$Department of Physics, University Sapienza, Piazzale Aldo Moro 5, 00185, Rome, Italy}
\affiliation{$^5$Department of Physical and Chemical Sciences, University of L'Aquila, Via Vetoio 10, 67100 L'Aquila, Italy}
\date{\today}
\begin{abstract}
We develop a novel theoretical framework describing polariton-enhanced spin-orbit interaction of light on the surface of two-dimensional media. Starting from the integral formulation of electromagnetic scattering, we exploit the reduced dimensionality of the system to introduce a quantum-like formalism particularly suitable to fully take advantage of rotational invariance. Our description is closely related to that of a fictitious spin one quantum particle living in the atomically thin medium, whose orbital, spin and total angular momenta play a key role in the scattering process. Conservation of total angular momentum upon scattering enables to physically unveil the interaction between radiation and the two-dimensional material along with the detailed exchange processes among orbital and spin components. In addition, we specialize our model to doped extended graphene, finding such spin-orbit interaction to be dramatically enhanced by the excitation of surface plasmon polaritons propagating radially along the graphene sheet. We provide several examples of the enormous possibilities offered by plasmon-enhanced spin-orbit interaction of light including vortex generation, mixing, and engineering of tunable deep subwavelength arrays of optical traps in the near field. Our results hold great potential for the development of nano-scaled quantum active elements and logic gates for the manipulation of hyper-entangled photon states as well as for the design of artificial media imprinted by engineered photonic lattices tweezing cold atoms into the desired patterns.
\end{abstract}
\maketitle

\section{Introduction}

Photons are light quanta characterized by energy and linear/angular momentum observables. While energy and linear momentum of photons are inherently linked to the wavelength, their angular momentum arises from both the spin and the phase pattern of the light wave, whose wavefront can be decomposed into orbital angular momentum (OAM) states of well defined topological charge \cite{Allen1992}. Standard protocols in quantum information are based on {\it qbits}, which in photonic realizations are composed of two-dimensional quantum states of photons with opposite spins arising from left and right polarizations ($|L\rangle$, $|R\rangle$) of the light wave. While the spin angular momenta of such states can take only values of $\pm \hbar$, the additional OAM of photons is generally unbound and can take any discrete value $l\hbar$, where $-\infty < l < +\infty$ is an integer accounting for the topological charge associated to the photon phase pattern. The resulting photon states -- composed by the coherent superposition of different polarizations and helical spatial patterns with definite topological charge -- live in a higher dimensional Hilbert space \cite{Mair2001,Molina2007,Nagali2010,Cardano2015} and enable the possibility to encode information into hyper-entangled states, also named {\it qdits}, which provide superior capacity for quantum cryptography \cite{Bechmann2000,Cerf2002,Mirho2015}. In addition to such promising quantum applications, the OAM of twisted light is currently employed for classical communications in the atmosphere \cite{Paterson2005,MalikOE2012,Willner2015,FarasSR2015,Goyal2016}, which are being demonstrated over increasing record-breaking distances \cite{TamburiniNJP2012,WangNP2012,Krenn2014,KrennPNAS2016}. Furthermore, the non-trivial interaction of twisted light with matter \cite{Pfeifer2007} plays a major role in laser cooling \cite{Chu1998,Phillips1998,Ashkin2000}, optical tweezing and manipulation \cite{Grier2003,Dienero2008,Bowman2013}, and many other applications where the spin-orbit interaction (SOI) of light \cite{MarrucciJOPT2011,Zayats2015} enables the controlling of the beam spatial pattern through its polarization.

SOI is an inherent property of the electromagnetic field that, similarly to relativistic quantum particles and electrons in solids \cite{Mathur1991,Xiao2010}, becomes relevant at spatial scales comparable with the optical wavelength \cite{Zhao2007} and is responsible for several unusual phenomena such as, for example, the spin-Hall effect of light \cite{Kavokin2005,Hosten2008,Ling2017,Bliokh2006}, Imbert-Fedorov shift \cite{Bliokh2013,Aiello2012} and vortex generation \cite{Ciattoni2003,Brasselet2009,Yavorsky2012,Ciattoni2017}. Owing to such inherent enhancement of SOI in the non-paraxial regime, focused beams or evanescently localized waves like surface plasmon polaritons (SPPs) \cite{OConnor2014,Pan2016} undergo several spin-dependent effects such as unidirectional propagation at interfaces \cite{Rodrig2013}. Graphene - an atomically-thin lattice of carbon atoms arranged in hexagonal pattern \cite{Geim2007} - offers unique possibilities for confining light down to the nanometer scale \cite{Bonaccorso2010}, with the further appealing ability of electrical tunability through gated injection of charge carriers. In particular, doped graphene enables the excitation of tunable surface plasmons in nano-islands and polaritons propagating along infinitely extended sheets \cite{JavierACSPhot} that play a relevant role in plasmon-enhanced light-matter interaction \cite{Koppens2011} leading to several applications in sensing \cite{Li2014,Rodrigo2015,Marini2015}, harmonic generation \cite{Mikhailov2011,Cox2017}, complete optical absorption \cite{Sukosin2012}, terahertz modulation \cite{Berardi2012,Berardi2013}, and many others. The extraordinary confinement offered by SPPs in extended graphene is expected to enhance dramatically the SOI of non-paraxial impinging light, leading to the development of active devices for the manipulation and control of OAM at the nanoscale.

\begin{figure*} \label{Fig1}
\center
\includegraphics[width=0.95\textwidth]{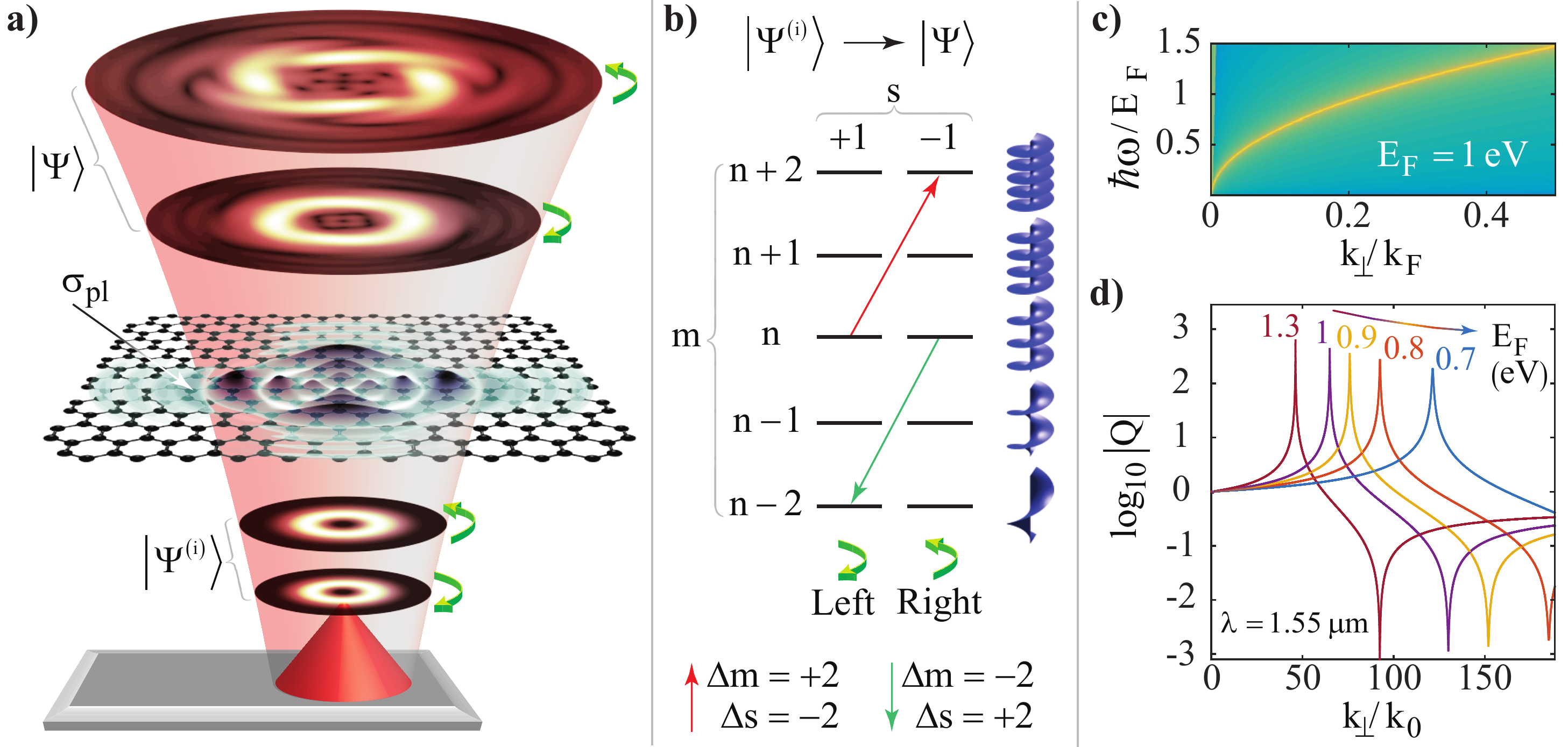}
\caption{(a) Geometry of radiation scattering by graphene. The impinging field (represented by the abstract vector $\left| {\Psi ^{\left( i \right)} } \right\rangle$) is radiated by a tip and it has both left and right handed polarized components. The circularly polarized components of the transmitted field (represented by the abstract vector $\left| \Psi  \right\rangle$) displays vortex photons not present in the incoming field. The high transverse confinement of the field emitted by the tip both enhances such vortex scattering process and enables the concurrent excitation of graphene surface plasmon polaritons whose charge oscillation is denoted with $\sigma_{\rm pl}$. (b) Vortex level scheme and fundamental scattering processes supported by graphene. A vortex $(m,s)$ is characterized by the topological charge $m$ (coinciding with its orbital angular momentum) and by the left/right circular polarization $s=\pm 1$ (coinciding with its spin). Graphene induces the transitions $(n,+1) \longrightarrow (n+2,-1)$ (red arrow) and $(n,-1) \longrightarrow (n-2,+1)$ (green arrow). The selection rules $\Delta m = \pm 2$ and $\Delta s = \mp 2$ are a consequence of the conservation of the total angular momentum $j=m+s$ (i.e. the rotational invariance of the system). (c) Illustration of the graphene plasmon dispersion relation $\omega(k_{\bot})$, where the photon energy $\hbar\omega$ is rescaled to the Fermi energy $E_{\rm F}= 1$ eV and $k_{\bot}$ is normalized to the Fermi wavevector $k_{\rm F}=E_{\rm F}/\hbar v_{\rm F} \simeq 1.5$ nm$^{-1}$. (d) Logarithmic plot of the vortex scattering function $|Q|$ introduced [See Eqs.(\ref{QP})] as a function of $k_\bot / k_0$ for $\lambda = 1.55 \: \mu$m at different Fermi energies from $0.7$ to $1.3$ eV. Each peak spectrally locates a graphene plasmon whose resonant transverse wavevectors $k_\bot$ strongly depend on the Fermi energy.}
\end{figure*}

Here we provide for the first time, to the best of our knowledge, a first-principle theoretical approach able to describe the scattering of arbitrary tightly confined fields impinging on a generic two-dimensional medium. Our framework, which does not resort to any approximation on the vectorial electromagnetic field, is very similar in its kinematical traits to the description of a spin one quantum particle living in the atomically thin medium, whose orbital, spin and total angular momenta are the basic quantities allowing us to elucidate the physics of radiation matter interaction in the presence of two-dimensional materials. Rotational invariance is the key ingredient of our analysis and accordingly we find that the ensuing total angular momentum conservation rules radiation scattering. Indeed, the rotationally invariant electromagnetic coupling provided by the two-dimensional material, triggers photon transitions accompanied by an exchange between the orbital and spin angular momenta which is quantified by suitable selection rules. We further specialize our calculations to doped extended graphene, demonstrating that thanks to plasmon-enhanced SOI such an atomically-thin medium can generate efficiently optical vortices of order $m = \pm 2$ and actively manipulate the mixing of different OAM states by means of the external gate voltage, thus enabling fast electrical processing of information stored in OAM states. In addition, we demonstrate that such a tunable mixing of OAM vortices of different order can be exploited to devise arbitrary subwavelength arrays of optical traps in the near field able to pin cold atoms into desired patterns, thus enabling the engineering of artificial materials at will. Such extraordinary functionalities of graphene ensue from the very large momentum of SPPs which, conversely to state-of-the-art optical tweezers, enable the generation of an interference pattern at deep subwavelength scales.

\section{Radiation scattering by two-dimensional media}

\subsection{Lippman-Schwinger equation}
Consider radiation scattering by a two-dimensional (2D) material lying on the plane $z=0$ embedded in vacuum and illuminated by a monochromatic radiation field (with time dependence $e^{-i \omega t}$, where $\omega$ is the angular frequency) impinging from the half-space $z<0$ (see Fig.1a). Owing to the reduced dimensionality of the system, the integral formalism is particularly suitable for describing radiation scattering. Exploiting the Green's function method, in the presence of a time harmonic current density of complex amplitude ${\bf J}({\bf r})$, the overall radiation electric field can be written as \cite{Schwinger}
\begin{eqnarray} \label{rad_field}
&& {\bf{E}}\left( {\bf{r}} \right) = {\bf{E}}^{\left( i \right)} \left( {\bf{r}} \right) + \nonumber \\ &+&  \frac{{i\omega \mu _0 }}{{4\pi }}\left( {1 + \frac{1}{{k_0^2 }}\nabla \nabla  \cdot } \right) \int {d^3 {\bf{r}}'} h\left( {{\bf{r}} - {\bf{r}}'} \right){\bf{J}}\left( {{\bf{r}}'} \right),
\end{eqnarray}
where $k_0 = \omega /c$, $c$ is the speed of light in vacuum and $\mu_0$ is the permeability of free space and $h\left( {\bf{r}} \right) = {e^{ik_0 r} } / r$ is the outgoing spherical wave; the field is the superposition of the incident field ${\bf{E}}^{\left( i \right)}$ (satisfying Maxwell equations in vacuum) and the field radiated by the current. In the limit of low impinging intensity, nonlinear effects are negligible and the current density induced by the external field on the atomically-thin medium is provided by
\begin{equation} \label{gr_cur_den}
{\bf{J}}\left( {\bf{r}} \right) = \sigma {\bf{E}}_ \bot  \left( {{\bf{r}}_ \bot} \right)\delta \left( z \right),
\end{equation}
where hereafter we set ${\bf{A}}_ \bot =  A_x{\bf{ e}}_x  + A_y{\bf{ e}}_y$ for the transverse part of a vector ${\bf{A}}$. Here $\sigma$ is the surface conductivity  of the 2D medium (see Sec. II.D for the particular case of graphene), which generally depends on the radiation frequency $\omega$ and the in-plane radiation wavevector ${\bf k}_{\bot}$. Inserting Eq.(\ref{gr_cur_den}) into Eq.(\ref{rad_field}) and separating transverse and longitudinal components we get
\begin{eqnarray} \label{scat}
&& {\bf{E}}_ \bot  \left( {\bf{r}} \right) = {\bf{E}}_ \bot ^{\left( i \right)} \left( {\bf{r}} \right) + \nonumber \\
 &+& \xi \left( {1 + \frac{1}{{k_0^2 }}\nabla _ \bot  \nabla _ \bot   \cdot } \right)\int {d^2 {\bf{r}}'_ \bot  } \frac{{ik_0 }}{{2\pi }} h \left( {{\bf{r}} - {\bf{r}}'_ \bot  } \right){\bf{E}}_ \bot  \left( {{\bf{r}}'_ \bot  } \right), \nonumber \\
&& E_z \left( {\bf{r}} \right) = E_z^{\left( i \right)} \left( {\bf{r}} \right) + \nonumber \\
 &+& \xi \left( {\frac{1}{{k_0^2 }}\frac{\partial }{{\partial z}}\nabla _ \bot   \cdot } \right)\int {d^2 {\bf{r}}'_ \bot  } \frac{{ik_0 }}{{2\pi }} h \left( {{\bf{r}} - {\bf{r}}'_ \bot  } \right){\bf{E}}_ \bot  \left( {{\bf{r}}'_ \bot  } \right),
\end{eqnarray}
where $\xi = \left( \mu_0 c/2 \right) \sigma$, and these equations are fully equivalent to differential Maxwell's equations supplemented
with the boundary conditions. Evaluating the first of Eqs.(\ref{scat}) at $z=0$ (i.e. setting ${\bf r} = {\bf r}_\bot$) we obtain the Lippman-Schwinger (LS) equation for ${\bf{E}}_ \bot  \left( {{\bf{r}}_ \bot } \right)$ describing radiation scattering from the 2D medium

\begin{equation} \label{lip-sch}
{\bf{E}}_ \bot  \left( {{\bf{r}}_ \bot  } \right) = {\bf{E}}_ \bot ^{\left( i \right)} \left( {{\bf{r}}_ \bot  } \right) + \xi V {\bf{E}}_ \bot  \left( {{\bf{r}}_ \bot  } \right),
\end{equation}
where the interaction operator is $V=MK$ with the operators $M$ and $K$ acting on transverse vectors ${\bf{A}}_ \bot$ according to
\begin{eqnarray}
 M{\bf{A}}_ \bot  & =& \left( {1 + \frac{1}{{k_0^2 }}\nabla _ \bot  \nabla _ \bot   \cdot } \right){\bf{A}}_ \bot,  \nonumber \\
 K{\bf{A}}_ \bot  & =& \int {d^2 {\bf{r}}'_ \bot  } \frac{{ik_0 }}{{2\pi }}h\left( {{\bf{r}}_ \bot   - {\bf{r}}'_ \bot  } \right){\bf{A}}_ \bot  \left( {{\bf{r}}'_ \bot  } \right).
\end{eqnarray}
Once the LS equation is solved for an incoming field ${\bf{E}}_ \bot ^{\left( i \right)} \left( {{\bf{r}}_ \bot } \right)$, Eqs.(\ref{scat}) provides the field ${\bf{E}}  \left( {{\bf{r}} } \right)$ in the whole space. It is worth noting that, conversely to standard bulk photonic media, it is possible to obtain an integral equation for the {\it single} transverse field ${\bf{E}}_ \bot  \left( {{\bf{r}}_ \bot } \right)$ owing to the in-plane surface current $\sigma {\bf{E}}_\bot$ flowing through the 2D medium. In particular, this enables us to perform the electromagnetic scattering analysis in the Hilbert space of square-integrable complex transverse vectors.

In order to shed light on the radiation SOI mechanism accompanying the scattering, it is convenient to exploit its rotational invariance around the $z$ axis: if $R_\vartheta$ is the rotation operator around the z-axis of an angle $\vartheta$ and if ${\bf{E}}_\bot$ is the field produced by the incoming field ${\bf{E}}_\bot^{(i)}$ than $R_\vartheta {\bf{E}}_\bot$ is the field produced by the incoming field $R_\vartheta {\bf{E}}_\bot^{(i)}$. From the LS equation it is straightforward to observe that the rotation and interaction operators commute, i.e. $[ R_\vartheta  ,V ] = 0$. Since we are dealing only with transverse vectors, the rotation operator is $R_\vartheta  = e^{ - iJ \vartheta }$ where $J =L +S$ and
\begin{eqnarray} \label{ang-mom-ope}
 L  &=&  \frac{1}{i} \left( {x\frac{\partial }{{\partial y}} - y\frac{\partial }{{\partial x}}} \right), \nonumber \\
 S  &=&  \frac{1}{i} \left( {\begin{array}{*{20}c}
   0 & 1  \\
   { - 1} & 0  \\
\end{array}} \right).
\end{eqnarray}
The operator $J$ is the infinitesimal generator of rotations around the $z$-axis so that, since other rotations are not involved here, by definition it is the total angular momentum operator and it is the sum of the orbital $L$ and spin $S$ angular momentum operators. For clarity, we note that the radiation angular momentum and its orbital and spin parts (which are vectors) should not be confused with $J$, $L$ and $S$ since the latter are purely geometrical operators suitably describing rotations around the z axis of transverse fields. Rotational invariance implies that total angular momentum and interaction operators commute, i.e. $[ J,V ] = 0$. This implies the conservation of angular momentum, i.e. the fact that if ${\bf{E}}_\bot^{(i)}$ is an eigenvector of $J$ then ${\bf{E}}_\bot$ is an eigenvector of $J$ as well with the same eigenvalue. Nevertheless, the separate orbital and spin angular momenta are not conserved, i.e. $[ L,V ] \neq 0$ and $[ S,V ] \neq 0$. This implies an effective {\it exchange} between them amounting to a {\it fundamental} SOI produced by the 2D medium, which fundamental physics is the main subject of this paper.

\subsection{Dirac's formalism and Vortex Representation}

The analysis of the physical mechanism supporting radiation scattering by a generic 2D medium is greatly simplified by using the Dirac's abstract vector space formalism. The Hilbert space which is suitable for our purposes is $\mathcal{H}=\mathcal{H}_{\rm orb} \otimes \mathcal{H}_{\rm spi}$,  the tensor product of the orbital and spin (polarization) state spaces  ${\cal H}_{\rm orb}  = \mathop {\rm span}\limits_{{\bf{r}}_ \bot  } \left\{ {\left| {{\bf{r}}_ \bot  } \right\rangle } \right\}$ and $\mathcal{H}_{\rm spi} = {\rm span} \left\{ {\left| {{\bf{e}}_x } \right\rangle ,\left| {{\bf{e}}_y } \right\rangle } \right\}$, respectively. Here ${\left| {{\bf{r}}_ \bot  } \right\rangle } $ are the eigenvectors of the transverse position operator and ${\left| {{\bf{e}}_x } \right\rangle ,\left| {{\bf{e}}_y } \right\rangle }$ are the abstract counterparts of the cartesian unit vectors. These two sets of vectors are orthonormal bases, i.e. $\int {d^2 {\bf{r}}_ \bot  } \left| {{\bf{r}}_ \bot  } \right\rangle \left\langle {{\bf{r}}_ \bot  } \right| = \hat{I}_{\rm orb}$, $\left\langle {{{\bf{r}}_ \bot  }}  \mathrel{\left | {\vphantom {{{\bf{r}}_ \bot  } {{\bf{r}}_ \bot  '}}} \right. \kern-\nulldelimiterspace} {{{\bf{r}}_ \bot  '}} \right\rangle  = \delta \left( {{\bf{r}}_ \bot   - {\bf{r}}_ \bot  '} \right)$ and ${\sum\nolimits_j {\left| {{\bf{e}}_j } \right\rangle \left\langle {{\bf{e}}_j } \right|}  = \hat I_{\rm spi} }$, $\left\langle {{{\bf{e}}_j }} \mathrel{\left | {\vphantom {{{\bf{e}}_j } {{\bf{e}}_{j'} }}} \right. \kern-\nulldelimiterspace} {{{\bf{e}}_{j'} }} \right\rangle  = \delta _{jj'}$, where $\hat{I}_{\rm orb}$ and $\hat{I}_{\rm spi}$ are the identity operators of $\mathcal{H}_{\rm orb}$ and $\mathcal{H}_{\rm spi}$, respectively (we hereafter use the caret $\wedge$ to label operators on abstract vector spaces). Accordingly the vectors $\left| {{\bf{r}}_ \bot  ,{\bf{e}}_j } \right\rangle  = \left| {{\bf{r}}_ \bot  } \right\rangle  \otimes \left| {{\bf{e}}_j } \right\rangle$ form an orthonormal basis of $\mathcal{H}$ and each transverse field ${\bf{A}}_ \bot   = A_x {\bf{e}}_x  + A_y {\bf{e}}_y$ is associated in $\mathcal{H}$ with the ket $\left| \Phi  \right\rangle  = \int {d^2 {\bf{r}}_ \bot  } \sum\nolimits_j {A_j \left( {{\bf{r}}_ \bot  } \right)\left| {{\bf{r}}_ \bot  ,{\bf{e}}_j } \right\rangle }$ with $A_j \left( {{\bf{r}}_ \bot  } \right) = \left\langle {{{\bf{r}}_ \bot  ,{\bf{e}}_j }} \mathrel{\left | {\vphantom {{{\bf{r}}_ \bot  ,{\bf{e}}_j } \Phi }} \right. } {\Phi } \right\rangle$.

The total angular momentum operator acting on $\mathcal{H}$ is $\hat J = \hat L + \hat S$  where, from Eqs.(\ref{ang-mom-ope}), the orbital and spin angular momentum operators are
\begin{eqnarray} \label{ang-mom-ope-abs}
 \hat L  &=& \frac{1}{i}\left( {\hat X\hat D_y  - \hat Y\hat D_x } \right) \otimes \hat I_{\rm spi}, \nonumber \\
 \hat S  &=& {\hat I}_{\rm orb}  \otimes \frac{1}{i}\left( {\left| {{\bf{e}}_x } \right\rangle \left\langle {{\bf{e}}_y } \right| - \left| {{\bf{e}}_y } \right\rangle \left\langle {{\bf{e}}_x } \right|} \right),
\end{eqnarray}
where $\hat X$, $\hat Y$ are the position operators (defined by $\left\langle {{\bf{r}}_ \bot  } \right|\hat X_i \left| {{\bf{r}}'_ \bot  } \right\rangle  = x_i \delta \left( {{\bf{r}}_ \bot   - {\bf{r}}'_ \bot  } \right)$) and $\hat D_x$, $\hat D_y$ are the derivatives operators (defined by $\left\langle {{\bf{r}}_ \bot  } \right|\hat D_i \left| {{\bf{r}}'_ \bot  } \right\rangle  = \partial _i \delta \left( {{\bf{r}}_ \bot   - {\bf{r}}'_ \bot  } \right)$). From the expression above it is straightforward to conclude that $\left[ {\hat L,\hat S} \right] = 0$.

In the space $\mathcal{H}$, the LS equation (Eq. \ref{lip-sch}) reads
\begin{equation} \label{lip-sch-abs}
\left| \Psi  \right\rangle  = \left| {\Psi ^{\left( i \right)} } \right\rangle  + \xi \hat V\left| \Psi  \right\rangle
\end{equation}
where $\left| {\Psi } \right\rangle$ and $\left| {\Psi ^{\left( i \right)} } \right\rangle$ are the ket associated with the field  ${\bf{E}}_\bot \left( {\bf r}_\bot \right)$ and the incoming field ${\bf{E}}_\bot^{(i)} \left( {\bf r}_\bot \right)$, respectively (see Fig.1a) and the interaction operator is $\hat V = \hat M \hat K$ with
\begin{eqnarray} \label{MK}
 \hat M &=& \hat{I} + \frac{1}{{k_0^2 }}\sum\limits_{j,j'} {\hat D_j \hat D_{j'}  \otimes \left| {{\bf{e}}_j } \right\rangle \left\langle {{\bf{e}}_{j'} } \right|}, \nonumber  \\
 \hat K &=& \int {d^2 {\bf{r}}_ \bot  } \int {d^2 {\bf{r}}'_ \bot  } \frac{{ik_0 }}{{2\pi }}h\left( {{\bf{r}}_ \bot   - {\bf{r}}'_ \bot  } \right)\left| {{\bf{r}}_ \bot  } \right\rangle \left\langle {{\bf{r}}'_ \bot  } \right|  \otimes \hat{I}_{\rm spi}, \nonumber \\
\end{eqnarray}
where $\hat{I}$ is the identity operator in $\mathcal{H}$.

The interaction operator $\hat V$ has a rather complicated structure which hampers the solution of the LS equation of Eq.(\ref{lip-sch-abs}). A convenient simplification is gained by resorting to a different representation whose basis vectors are more suitable to deal with the rotational invariance of the system. The basic observation is that the operator $\hat K$ commutes with both the orbital and spin angular momentum operators (mainly since it produces the convolution with the spherical wave and it does not affect the polarization of the field) so that $\hat K, \hat L , \hat S$ are a complete set of commuting operators in $\mathcal{H}$ and accordingly their common eigenvectors turn out to form an orthonormal basis of $\mathcal{H}$. In Appendix \ref{APP-eigenvectors} we show that these eigenvectors are
\begin{eqnarray} \label{vortex}
\left| {k_ \bot  ,m,s} \right\rangle  &=& \int {d^2 {\bf{r}}_ \bot  } \sqrt {\frac{{k_ \bot  }}{{2\pi }}} J_m \left( {k_ \bot  r_ \bot  } \right)e^{im\varphi } \left| {{\bf{r}}_ \bot  } \right\rangle \otimes \nonumber \\
  &\otimes& \frac{1}{{\sqrt 2 }}\left( {\left| {{\bf{e}}_x } \right\rangle  + is\left| {{\bf{e}}_y } \right\rangle } \right)
\end{eqnarray}
where $k_ \bot$ is any real and positive number, $m$ is an arbitrary integer and $s=\pm 1$. We also prove they satisfy the relations
\begin{eqnarray}
 \left\langle {k_ \bot  ,m,s} \right.|\left. {k'_ \bot  ,m',s'} \right\rangle  &=& \delta \left( {k_ \bot   - k'_ \bot  } \right)\delta _{mm'} \delta _{ss'}, \nonumber  \\
 \sum\limits_{k_ \bot  ,m,s} {|\left. {k_ \bot  ,m,s} \right\rangle \left\langle {k_ \bot  ,m,s} \right.|\;}  &=& \hat I,
\end{eqnarray}
where, according to Eq.(\ref{shorthand}) of Appendix \ref{APP-eigenvectors}, $\sum\limits_{k_ \bot  ,m,s}$ is a shorthand for the integration over $k_ \bot$ and the sum over $m$ and $s$. By construction, we have
\begin{eqnarray} \label{eigen}
 \hat K\left| {k_ \bot  ,m,s} \right\rangle  &=& u \left( k_\bot \right) \left| {k_ \bot  ,m,s} \right\rangle, \nonumber  \\
 \hat L\left| {k_ \bot  ,m,s} \right\rangle  &=& m \left| {k_ \bot  ,m,s} \right\rangle, \nonumber  \\
 \hat S\left| {k_ \bot  ,m,s} \right\rangle  &=& s \left| {k_ \bot  ,m,s} \right\rangle,
\end{eqnarray}
where $u \left( {k_ \bot  } \right) =  - \left( {1 - k_ \bot ^2 /k_0^2 } \right)^{ - 1/2}$. From the electromagnetic point of view the field associated with $\left| {k_ \bot  ,m,s} \right\rangle$ has the spatial profile $\sim J_m \left( {k_ \bot  r_ \bot  } \right)e^{im\varphi }$, i.e. it is has a Bessel profile of order $m$ endowed with a vortex of topological charge $m$ and it is left and right handed circularly polarized for $s=+1$ and $s=-1$, respectively, since these two values are associated to the unit vectors ${\bf{e}}_L  = \frac{1}{\sqrt 2} \left( {{\bf{e}}_x  + i{\bf{e}}_y } \right)$, ${\bf{e}}_R  = \frac{1}{\sqrt 2} \left( {{\bf{e}}_x  - i{\bf{e}}_y } \right)$, respectively. Hence $k_\bot$ is the radial transverse momentum, i.e. it is the radius of the Bessel ring in the momentum space. The second and the third of Eqs.(\ref{eigen}) imply that $m$ and $s$ are the orbital and spin angular momenta of the state $\left| {k_ \bot  ,m,s} \right\rangle$. This allows to unambiguously identify the orbital angular momentum with the vortex topological charge and the spin with the vortex polarization. Again, we emphasize that such angular momenta {\it are not} the standard electromagnetic ones but are the eigenvalues of the infinitesimal generators of rotations in the orbital and spin state spaces. We hereafter use the basis $\left| {k_ \bot  ,m,s} \right\rangle$, which we name the {\it vortex basis}, since it provides the most suitable representation for our purposes (the {\it vortex representation}).

Owing to the rotational symmetry, the vortex $\left| {k_ \bot  ,m,s} \right\rangle$ is an eigenvector of the total angular momentum $\hat J = \hat L + \hat S$
\begin{equation}
\hat J|\left. {k_\bot,m,s} \right\rangle  = \left( {m + s} \right)|\left. {k_\bot,m,s} \right\rangle
\end{equation}
with eigenvalue $j=m+s$. Therefore each eigenvalue of $\hat J$ is two-fold degenerate and its two-dimensional eigenspace is
\begin{equation} \label{E(kj)}
\mathcal{E} \left( {k_ \bot  ,j} \right) = {\rm span} \left\{ {|\left. {k_\bot,j - 1, + 1} \right\rangle ,\;|\left. {k_\bot ,j + 1, - 1} \right\rangle } \right\}.
\end{equation}
Since the vortex basis is orthonormal, it follows that the Hilbert space $\mathcal{H}$ is the direct sum of the eigenspaces $\mathcal{E} \left( {k_ \bot  ,j} \right)$ [i.e. $\mathcal{H} =  \bigoplus _{k_ \bot  ,j} \mathcal{E}\left( {k_ \bot  ,j} \right)$].

A geometrical characterization of the vortex representation is gained by exploiting the fact that here only rotations in the plane are allowed and hence the symmetry group is $SO(2)$ which, being Abelian, has only one-dimensional irreducible representation. In Appendix \ref{APP-eigenvectors}, we prove that each vector $|\left. {k_ \bot  ,m,s} \right\rangle$ is the basis of an irreducible representation of $SO(2)$.

\subsection{Fundamental radiation SOI processes} \label{SOI}
The vortex representation introduced in the above section, besides providing a framework for associating the topological charge with the orbital angular momentum and the circular polarizations with the spin states, is also very useful for handling the interaction of light with 2D media. To achieve this goal, let us consider the operators
\begin{equation}
\begin{array}{*{20}c}
   {\hat L_ +   = \left( {\hat D_x  + i\hat D_y } \right) \otimes \hat I_{\rm spi} ,} & \quad {\hat S_ +   = \hat I_{\rm orb}  \otimes \left| { + 1} \right\rangle \left\langle { - 1} \right|,}  \\
   {\hat L_ -   = \left( {\hat D_x  - i\hat D_y } \right) \otimes \hat I_{\rm spi} ,} & \quad {\hat S_ -   = \hat I_{\rm orb}  \otimes \left| { - 1} \right\rangle \left\langle { + 1} \right|,}  \\
\end{array}
\end{equation}
which, as shown in Appendix \ref{APP-ladder}, are such that
\begin{eqnarray} \label{ladder-ops}
 \hat L_ \pm  \left| {k_ \bot  ,m,s} \right\rangle  &=&  \mp k_ \bot  \left| {k_ \bot  ,m \pm 1,s} \right\rangle, \nonumber \\
 \hat S_ \pm  |\left. {k_ \bot  ,m,s} \right\rangle  &=& \delta _{s, \mp 1} |\left. {k_ \bot  ,m,s \pm 2} \right\rangle,
\end{eqnarray}
so that $\hat L_\pm$ and $\hat S_\pm$ are orbital and spin ladder operators (the former raising and lowering $m$ by one unit and the latter raising and lowering $s$ by two units). The crucial point, that fully justifies our choice to use the vortex representation, is that the operators $\hat M$ and $\hat K$ of Eqs.(\ref{MK}) can be expressed only in terms of the ladder operators as
\begin{eqnarray}
\hat M &=& \hat I + \frac{1}{{2k_0^2 }} \left( \hat L_ +  \hat L_ - +\hat L_ + ^2 \hat S_ -   + \hat L_ - ^2 \hat S_ +   \right), \nonumber \\
 \hat K &=&  - \left( {\hat I + \frac{1}{{k_0^2 }}\hat L_ +  \hat L_ -  } \right)^{ - 1/2},
\end{eqnarray}
and these expressions are very compact and sufficiently simple to unveil the physical mechanisms supporting radiation scattering from the 2D medium (see Appendix \ref{APP-inter}). Note that in Appendix \ref{APP-inter} we also prove (as anticipated above) that both $\hat M$ and $\hat K$ are rotationally invariant operators (i.e. $[\hat M, \hat J] = [\hat K, \hat J] = 0$) so that the interaction operator $\hat V = \hat M \hat K$ is also a rotationally invariant operator ($[\hat V, \hat J]=0$). In particular, this implies that the eigenspaces $\mathcal{E} \left( {k_ \bot  ,j} \right)$ are invariant for the interaction operator $\hat V$ so that the total angular momentum $\hat J$ is conserved in the scattering process. However, as opposed to $\hat K$, the operator $\hat M$ conserves neither the orbital nor the spin angular momentum so that the interaction operator $\hat V$ will in general trigger transitions among states of different $m$ and $s$ while conserving their sum $j=m+s$.

Before discussing the general solution of the LS equation [see Eq.(\ref{lip-sch-abs})], we first examine its solution in the Born approximation (i.e. up to the first order in $\xi$), namely
\begin{equation} \label{LS-Sol-app}
\left| \Psi  \right\rangle  = \left( {\hat I + \xi \hat M\hat K} \right)\left| {\Psi ^{\left( i \right)} } \right\rangle.
\end{equation}
If the incoming field is a vector of the vortex basis, $\left| {\Psi ^{\left( i \right)} } \right\rangle  = \left| {k_ \bot  ,m,s} \right\rangle$, by using Eqs.(\ref{ladder-ops}), we obtain
\begin{eqnarray}
\left| \Psi  \right\rangle  &=& \left[ {1 + \xi \left( {1 - \frac{{k_ \bot ^2 }}{{2k_0^2 }}} \right)u} \right] \left| {k_ \bot  ,m,s} \right\rangle  + \nonumber \\
 &-& \left( \xi \frac{{k_ \bot ^2 }}{{2k_0^2 }}u \right) \delta _{s, + 1} \left| {k_ \bot  ,m + 2,s - 2} \right\rangle + \nonumber \\
 &-& \left( \xi \frac{{k_ \bot ^2 }}{{2k_0^2 }}u \right) \delta _{s, - 1} \left| {k_ \bot  ,m - 2,s + 2} \right\rangle.
\end{eqnarray}
Therefore the 2D medium triggers transitions from a vortex of orbital momentum $m$ and spin $s=+1$ to one of angular momentum $m+2$ and spin $s-2=-1$ (due to the interaction operator $\hat L_ + ^2 \hat S_ -$) and from a vortex of orbital momentum $m$ and spin $s=-1$ to one of angular momentum $m-2$ and spin $s+2=+1$ (due to the interaction operator $\hat L_ - ^2 \hat S_ +$). These two scattering processes (see Fig.1b) are very fundamental (see below) and they provide the physical background for understanding all the photonic spin-orbit interaction produced by 2D media. Note that the total angular momentum $j=m+s$ is always conserved in the two processes, as expected due to the above discussed rotation invariance of the system. Total angular momentum conservation also explains the selection rule $\Delta m = \pm 2$ since in each process the orbital angular momentum change has to compensate the change $\Delta s = \mp 2$ accompanying the flip of the spin (which has only the eigenvalues $+1$ and $-1$, see Fig.1b). Translated into the electromagnetic language, if we illuminate the 2D material with a left (right) handed circularly polarized vortex of topological charge $m$ the scattered field will also contain a right (left) handed circularly polarized vortex of topological charge $m+2$ ($m-2$). Note that the scattering amplitude $\xi \frac{{k_ \bot ^2 }}{{2k_0^2 }}u\left( {k_ \bot  } \right)$ reveals that by increasing the ratio $k_\bot / k_0$ the transitions' efficiency increases as well: the more transversely confined the incoming field is, the more strong are the scattered vortices.

The general solution of the LS equation (Eq.(\ref{lip-sch-abs})) is (see Appendix \ref{APP-LS})
\begin{eqnarray} \label{LS-Sol}
 \left| \Psi  \right\rangle  &=& \sum\limits_{k_ \bot  ,m,s} {\psi _{k_ \bot  ,m,s}^{\left( i \right)} Q\left( {k_ \bot  } \right)|\left. {k_ \bot  ,m,s} \right\rangle }  + \nonumber \\
  &+& \sum\limits_{k_ \bot  ,m,s} {\psi _{k_ \bot  ,m,s}^{\left( i \right)} P\left( {k_ \bot  } \right)|\left. {k_ \bot  ,m + 2s, - s} \right\rangle }
\end{eqnarray}
where $\psi _{k_ \bot  ,m,s}^{\left( i \right)}  = \left\langle {{k_ \bot  ,m,s}}  \mathrel{\left | {\vphantom {{k_ \bot  ,m,s} {\Psi ^{\left( i \right)} }}} \right. \kern-\nulldelimiterspace} {{\Psi ^{\left( i \right)} }} \right\rangle$ are the Fourier coefficients of the expansion of the incoming state in the vortex basis and
\begin{eqnarray} \label{QP}
 Q &=& \frac{1}{2}\left( {\frac{1}{{1 - \xi u}} + \frac{1}{{1 - \xi u^{ - 1} }}} \right), \nonumber \\
 P &=& \frac{1}{2}\left( {\frac{1}{{1 - \xi u}} - \frac{1}{{1 - \xi u^{ - 1} }}} \right). \label{GRESEQ}
\end{eqnarray}
Note that the first term in the RHS of Eq.(\ref{LS-Sol}) has the same vortex structure of the incoming state whereas the second term describes the very same scattering processes $\left( {k_ \bot  ,m, + 1} \right) \to \left( {k_ \bot  ,m + 2, - 1} \right)$ and $\left( {k_ \bot  ,m, - 1} \right) \to \left( {k_ \bot  ,m - 2, + 1} \right)$ described above in the Born approximation. Conversely to Eq.(\ref{LS-Sol-app}), the solution of Eq.(\ref{LS-Sol}) is non-perturbative and accounts for transverse magnetic (TM) and transverse electric (TE) resonances provided by the conditions ${\rm Re}\left(1 - \xi u\right)=0$ and ${\rm Re}\left(1 - \xi u^{-1}\right)=0$, respectively.

\subsection{Graphene Plasmonic Resonance}

The novel theoretical treatment of SOI of light discussed above is general for any kind of 2D material, which physical properties are fully incorporated within the parameter $\xi$ depending on the surface conductivity $\sigma$. In the particular case of graphene, at optical and infrared frequencies the response is dominated by the conical band structure ${\cal E} = \pm v_{\rm F} p$ around the two Dirac points of the first Brillouin zone, where $v_{\rm F} \approx 10^6$ m$/$s is the Fermi velocity and ${\cal E},{\bf p}$ are the electron energy and momentum, respectively \cite{CastroNeto2009}. While in undoped graphene the Fermi energy lies at the Dirac points, injection of charge carriers through electrical gating \cite{Chen2011} or chemical doping \cite{Liu2011} shifts efficiently the Fermi level up to $E_{\rm F} \approx 1-2$ eV owing to the conical dispersion and the 2D electron confinement. Thus, within the photon energy range $\hbar \omega < 2 E_{\rm F}$, where $\hbar$ is the reduced Planck constant, interband transitions are inhibited by the Pauli exclusion principle and graphene acquires a metal-like behavior \cite{Bonaccorso2010}. In turn, doping affects enormously the optical response of graphene from an efficient dispersionless absorber of $\approx 2.3 \%$ of impinging radiation (and universal conductivity $\sigma_0 = e^2/4\hbar$, where $-e$ is the electron charge) to a 2D metal with long relaxation time $\tau = \mu E_{\rm F}/ev_{\rm F}^2$, where $\mu$ is the electron mobility, which conversely to noble metals can reach the picosecond time scale at moderate doping and purity (affecting electron mobility) \cite{JavierACSPhot}. In principle the graphene surface conductivity $\sigma({\bf k}_{\bot},\omega)$ is nonlocal and depends on the in-plane radiation wavevector ${\bf k}_{\bot}$. However, when $k_{\bot}<k_{\rm F} = E_{\rm F}/\hbar v_{\rm F}$ electron dynamics is local and the graphene conductivity can be evaluated within the local random phase approximation (RPA) providing the integral expression
\begin{eqnarray}
\sigma(\omega) & = & \frac{-ie^2}{\pi\hbar^2(\omega+i/\tau)}\int_{-\infty}^{+\infty}d{\cal E} \left\{|{\cal E}| \frac{ \partial f_{\cal E} }{\partial {\cal E}} + \right. \nonumber \\
               &   & \left. + \frac{ f_{\cal E} {\cal E}/|{\cal E}| }{1 - 4 {\cal E}^2/[\hbar(\omega+i/\tau)]^2} \right\},
\end{eqnarray}
where $f_{\cal E} = \{ 1 + {\rm exp}[-({\cal E} - E_{\rm F})/k_{\rm B} T]\}$ is the Fermi-Dirac occupation density of states up to the Fermi energy $E_{\rm F}$, $k_{\rm B}$ is the Boltzmann constant and $T$ is the electron temperature.

Such metal-like behavior of doped graphene enables the tight coupling of photons with in-plane plasma oscillations (pictorially depicted in Fig. 1a as $\sigma_{\rm pl}$, leading to the existence of exponentially localized TM electromagnetic modes propagating along the infinitely extended sheet with wavevector ${\bf k}_{\bot}$ \cite{JavierACSPhot}. The dispersion relation $\omega(k_{\bot})$ of such modes is identified exactly by the quasi-pole ${\rm Re}\left(1 - \xi K\right)=0$ derived in Eq. (\ref{GRESEQ}) and is depicted in Fig.1c, from which one can observe that graphene can confine SPPs down to approximately $10$ nm before non-local effects come into play counteracting further localization. Such an extraordinary confinement at the deep subwavelength nanoscale provided by graphene SPPs enables the enhancement of SOI of light by several orders of magnitude, as we discuss below.

For monochromatic fields, the graphene surface plasmon resonance occurs at a specific transverse wavevector $k_\bot$ which strongly depends on the Fermi energy of the sheet. In Fig.1d, we set $\lambda = 1.55 \: \mu$m and we depict the logarithmic plot of $|Q|$ introduced in the first of Eqs.(\ref{QP}) as a function of $k_\bot / k_0$ for different Fermi energies from $0.7$ to $1.3$ eV. The peaks identify the spectral position of the graphene SPPs, whose transverse wavevectors $k_\bot$ shift from $50 \: k_0$ to $120 \: k_0$ for different Fermi energies.

\section{Vortex Management}

\subsection{Electromagnetic vortex scattering}
The above discussed theory on radiation scattering by graphene can be profitably exploited, as we are going to prove, in a number of relevant applications. As depicted in Fig.1a, the general scheme deals with illuminating the graphene sheet with an optical radiation whose circularly polarized components contain suitable vortices which are scattered into novel vortices setting up a desired profile of the transmitted field. As discussed in subsection \ref{SOI}, the vortex scattering efficiency is very small if the impinging field has a transverse confinement much greater than the wavelength. Besides, in order to make efficient the radiation-plasmon coupling and hence to take advantage of the associated resonance, the evanescent content of the incoming field necessarily has to be large. Both these conditions (which are essentially equivalent) can be achieved by using a near-field optical probe (the tip in Fig.1a) to generate the incoming field.

\begin{figure*}
\center
\includegraphics[width=0.95\textwidth]{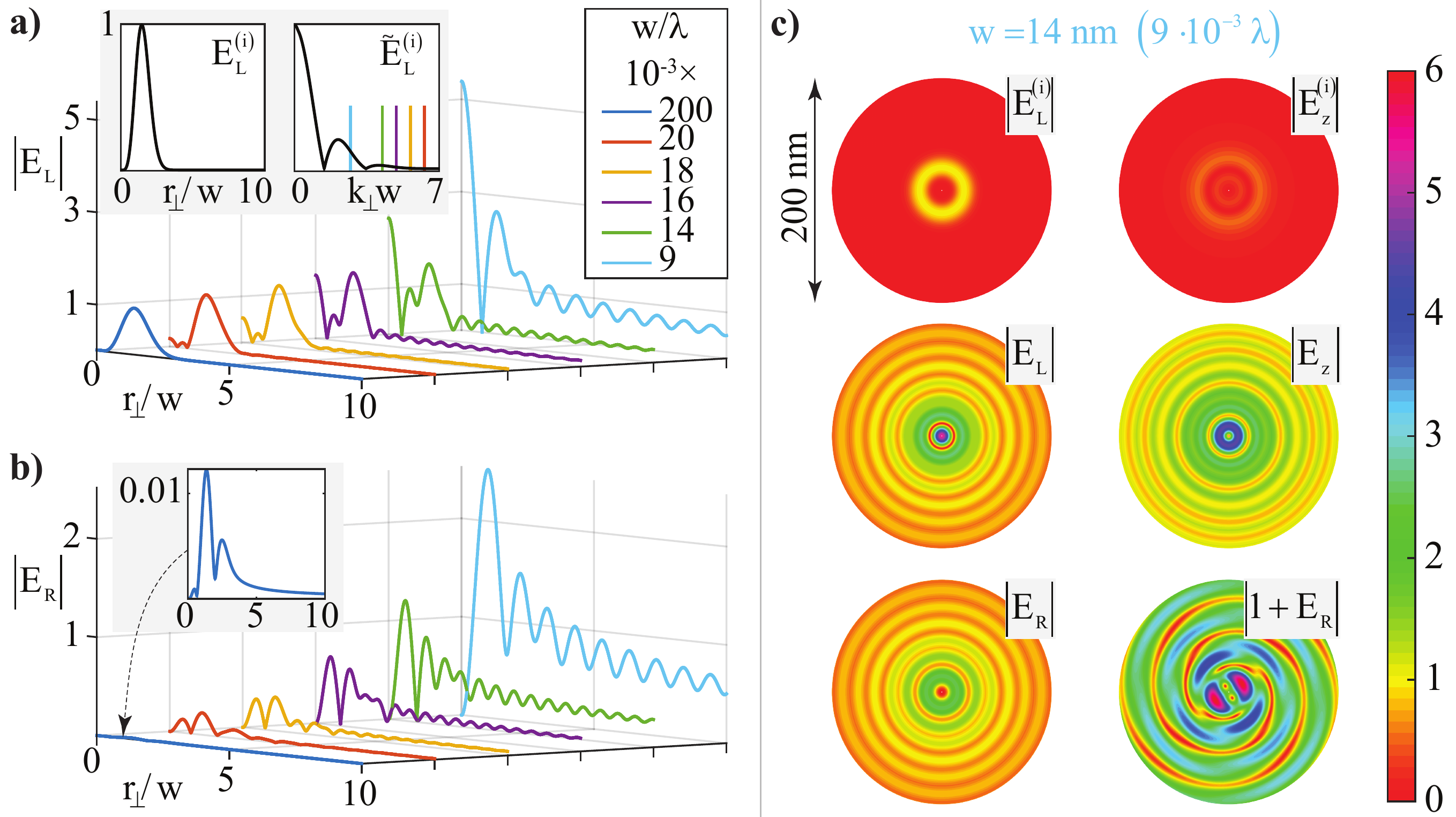}
\caption{Vortex generation at the wavelength $\lambda = 1.55 \: \mu$m from a graphene sheet whose Fermi energy is $1.14$ eV. In the inset of (a) the profile $E^{(i)}_L$ of the vortex-free left hand circularly polarized incoming field is plotted, together with its Hankel transform $\tilde{E}^{(i)}_L$. (a,b) Profiles of the circular components of the scattered field $|E_L|$ and $|E_R|$ are plotted for six different values of the incoming field width $w$. The component $E_R$ is endowed with a topological charge two and the corresponding vortex generation efficiency is increasingly higher for smaller widths $w$. The ripple appearing in the scattered field profile, which is increasingly stronger for smaller widths $w$, is the signature that a radial surface plasmon polariton has been excited with a concomitant plasmonic resonance. For each of the last five values of $w$, a vertical segments of corresponding color is plotted in the right inset of (a) to spectrally locate the corresponding plasmonic resonance (occurring at $k_\bot \simeq 54 k_0$) on the $k_\bot w$ axis and to check that the incoming field can actually couple with the surface plasmon polariton. (c) Two-dimensional plots of field amplitudes for $w=9 \cdot 10^{-3} \: \lambda = 14$ nm. In addition to the $|E_L^{(i)}|$, $|E_L|$ and $|E_R|$ corresponding to those of (a) and (b) we plot the amplitudes of the longitudinal components $|E_z^{(i)}|$ and $|E_z|$ of the incoming and scattered field, whose comparison reveals a remarkable field enhancement due to the plasmonic resonance. We also plot the amplitude $|1+E_R|$, which corresponds to the superposition of the generated vortex with an additional right hand circularly polarized plane wave, whose interference spiral shape is a signature of the vortex topological charge carried by $E_R$.}
\end{figure*}

In order to illustrate the phenomenology of the radiation SOI and to discuss its relevant applications, we evaluate the transmitted field ${\bf{E}} = E_L {\bf{e}}_L  + E_R {\bf{e}}_L  + E_z {\bf{e}}_z$ at $z=0^+$ in turn produced by the incoming field ${\bf E}^{(i)}_\bot \left( \bf{r}_\bot \right)$. We hereafter use circularly polarized components to describe the transverse part of the field which, in turn, satisfies the LS equation (due to the continuity of tangential component of the electric field across the graphene plane) so that $E_L \left( {{\bf{r}}_ \bot  } \right) = \left\langle {{{\bf{r}}_ \bot  , + 1}}  \mathrel{\left | {\vphantom {{{\bf{r}}_ \bot  , + 1} \Psi }} \right. \kern-\nulldelimiterspace} {\Psi } \right\rangle$ and $E_R \left( {{\bf{r}}_ \bot  } \right) = \left\langle {{{\bf{r}}_ \bot  , - 1}} \mathrel{\left | {\vphantom {{{\bf{r}}_ \bot  , - 1} \Psi }} \right. \kern-\nulldelimiterspace}  {\Psi } \right\rangle$, where $\left| {{\bf{r}}_ \bot  ,s} \right\rangle  = \left| {{\bf{r}}_ \bot  } \right\rangle  \otimes \left| s \right\rangle$. The longitudinal component $E_z$ can be evaluated both from the second of Eqs.(\ref{scat}) or from the divergence-free property of the electric field in vacuum (see Appendix \ref{APP-field}). Here such longitudinal component can not be neglected since, as above anticipated, we are going to discuss near-field optical applications where the fields are highly confined in the transverse plane and plasmonic resonance generally plays a key role. In addition, in order to discuss the spatial energy redistribution accompanying graphene SOI of light, we also consider the time-averaged Poynting vector pertaining the transmitted field at $z=0^+$, i.e. ${\bf{S}} = \frac{1}{2}{\mathop{\rm Re}\nolimits} \left( {{\bf{E}} \times {\bf{H}}^* } \right)$ (see Appendix \ref{APP-field} for the evaluation of the magnetic field ${\bf{H}}$).

\subsection{Vortex Generation} \label{VG}
As a first application of the SOI of light produced by the graphene sheet, we discuss the generation of optical vortices out of an incoming field with vanishing topological charge. Consider an impinging monochromatic field of wavelength $\lambda = 1.55 \: \mu$m whose  profile at the graphene sample (with Fermi energy $E_{\rm F} = 1.14$ eV) is assumed

\begin{eqnarray}
 {\bf{E}}_ \bot ^{\left( i \right)} \left( {{\bf{r}}_ \bot  } \right) &=& E_L^{\left( i \right)} \left( {r_ \bot  } \right){\bf{e}}_L  = \nonumber  \\
  &=& E_0 \frac{{e^2 }}{4}\left( {\frac{{r_ \bot  }}{w}} \right)^4 e^{ - \left( {\frac{{r_ \bot  }}{w}} \right)^2 } {\bf{e}}_L.
\end{eqnarray}
This is a left hand circularly polarized field with vanishing topological charge (i.e. it has no vortex singularity at $\bf{r}_ \bot = \bf{0}$.) and the profile of $E_L^{\left( i \right)}$ is depicted in the left inset of Fig.2a. Apart from the amplitude $E_0$ (which we hereafter set equal to one), its only feature which is relevant for our discussion is the width $w$. Such impinging field is the superposition of only the basis vortices $(m,s)=(0,+1)$ so that its total angular momentum is $j=+1$ and its spectral content is provided by the Hankel transform of order zero (see Eq.(\ref{psi}))
\begin{equation}
\tilde E_L^{\left( i \right)} \left( {k_ \bot  } \right) = \int\limits_0^{ + \infty } {dr_ \bot  r_ \bot  J_0 \left( {k_ \bot  r_ \bot  } \right)} E_L^{\left( i \right)} \left( {r_ \bot  } \right),
\end{equation}
which is depicted in the right inset of Fig.2a.

Due to SOI of light, the field scattered by the graphene sheet has the structure
\begin{eqnarray} \label{scatVort}
 E_L \left( {{\bf{r}}_ \bot  } \right) &=& E_{LL} \left( {r_ \bot  } \right), \nonumber \\
 E_R \left( {{\bf{r}}_ \bot  } \right) &=& E_{RL} \left( {r_ \bot  } \right)e^{i2\varphi }, \nonumber \\
 E_z \left( {{\bf{r}}_ \bot  } \right) &=& E_{zL} \left( {r_ \bot  } \right)e^{i\varphi },
\end{eqnarray}
whose amplitudes have close form integral expressions (see Appendix \ref{APP-field}) which can be numerically evaluated. This scattered field has both left and right hand circularly polarized components, the former with circular symmetry and the latter with topological charge two (both with the total angular momentum $j=+1$). In other words, a vortex of topological charge two has been generated by the graphene sheet. The longitudinal component has topological charge $1$ (as a consequence of the divergence-free property of the electric field in vacuum) and, if we associate the spin value $s=0$ to $E_z$, it also has total angular momentum $j=+1$.

\begin{figure}
\center
\includegraphics[width=0.45\textwidth]{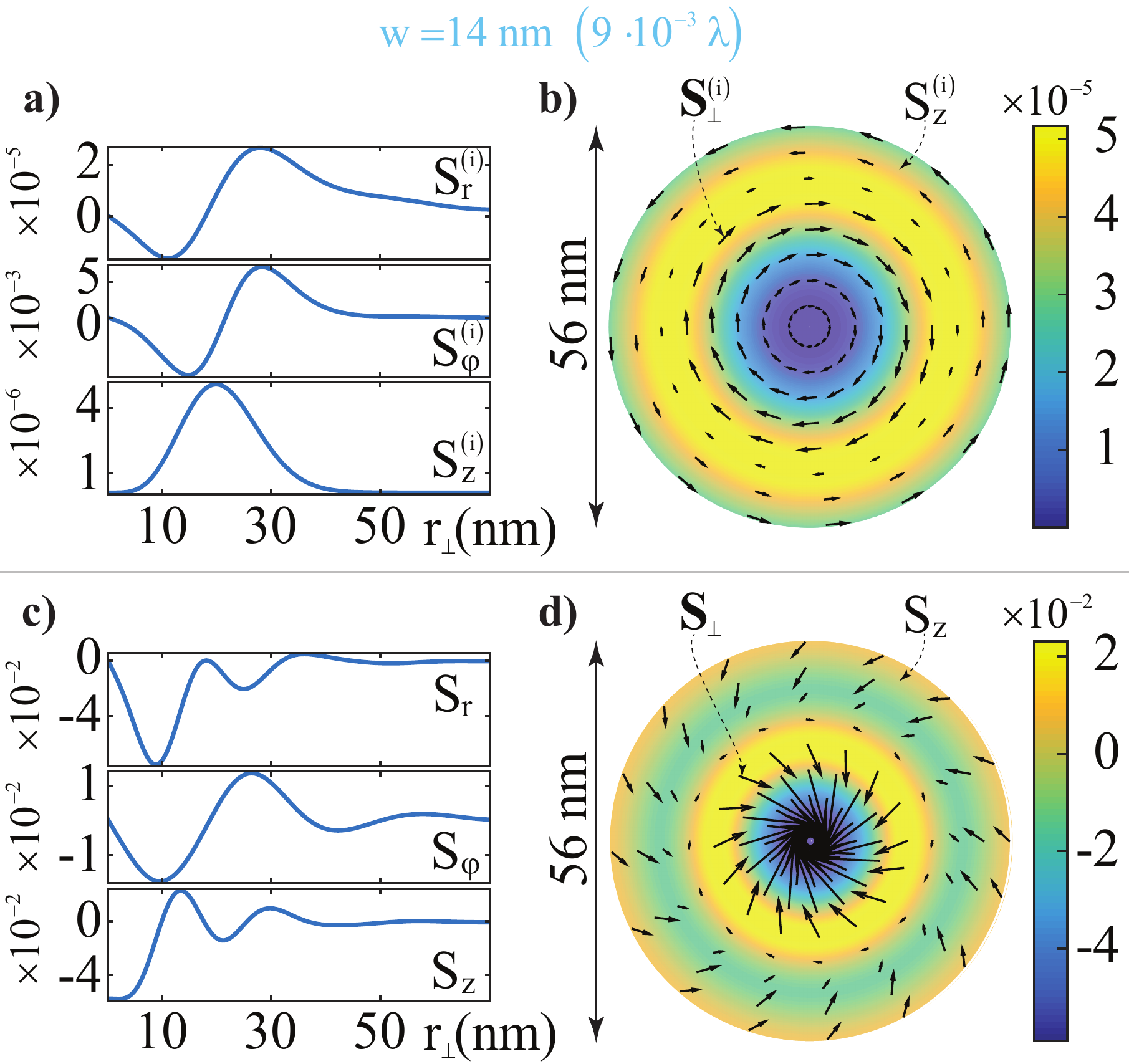}
\caption{Energy flow redistribution accompanying the vortex generation process discussed in Fig.2c ($w=9 \cdot 10^{-3} \: \lambda = 14$ nm). (a,c) Cylindrical components of the incoming and scattered Poynting vectors, respectively. (b,d) Pictorial representation of the transverse part (as a vector field) and of the longitudinal component (as a color plot) of the incoming and scattered Poynting vectors, respectively. The interplay between graphene SOI of light and plasmon resonance produces a consistent enhancement of all the Poynting vector components resulting in a spatial energy flow pattern very much different from the purely azimuthal circulating one pertaining the incident field.}
\end{figure}

In Figs.2a and 2b we plot the profiles of $|E_L|$ and $|E_R|$, respectively, for six different values of the incoming field width, i.e. $w = 200, 20, 18, 16, 14, 9 \cdot 10^{-3} \lambda$. Note from Fig.2b that the amplitude of the generated vortex changes from $\simeq 0.01$ to $\simeq 2.5$ when $w$ is reduced from $200 \cdot 10^{-3} \lambda$ to $9 \cdot 10^{-3} \lambda$. The fact that the efficiency of the vortex generation is increasingly higher for smaller widths $w$ is a general trait of SOI of light produced by graphene, as discussed in Sec. \ref{SOI}. A second remarkable feature which is evident from Figs.2a and 2b is that, as the width $w$ is decreased, an increasingly stronger ripple appears on the radial profiles of both the circularly polarized components $E_L$ and $E_R$. Such a ripple is the signature of the coupling between the impinging field and the surface plasma oscillations of the graphene sheet, coupling resulting into the excitation of a radial SPP accompanied by a strong plasmon resonance. Since the chosen Fermi energy is $1.14$ eV, such plasmon resonance occurs at $k_\bot \simeq 54 \: k_0$ (see Fig.1d) and in the right inset of Fig.2a we have denoted the spectral position of the resonances $k_\bot w$ by vertical segments whose colors correspond to the values of $w$ considered in the simulations. Evidently, the smaller $w$, the closer the plasmonic resonance $k_\bot w$ to the region where the field spectrum has it central lobe. In other words, the more the field is tightly confined, the more strong is its evanescent spectral tail overlapping the plasmon resonance. Note also that, in addition to the radial ripple, the plasmonic resonance also increases the efficiency of the SOI of light produced by graphene. This is particularly evident from Fig.2a since the amplitude of the left hand component $E_L$ is practically equal to $1$ when the width $w$ is greater than $18 \cdot 10^{-3} \: \lambda$ (yellow curve) and it raises to $\sim 5$ for $w = 9 \cdot 10^{-3} \: \lambda$.

In Fig.2c we focus on the considered most tight confined field  with $w=9 \cdot 10^{-3} \: \lambda = 14$ nm and we discuss some of its spatial features on the transverse plane. For comparison purposes, all the amplitudes are plotted on a disk centered at ${\bf r}_\bot = {\bf 0}$ and diameter of $200$ nm with the same color scale reported on right side. The amplitudes $|E_L^{(i)}|$, $|E_L|$ and $|E_R|$ are the two-dimensional counterparts of those considered in Figs.2a and 2b. Note that the scattered field has a multiple ring structure with a spatial extension much wider that the incoming field and this is a consequence of the large radial SPP decay length. The comparison between the longitudinal components of the incoming and scattered field, $|E_z^{(i)}|$ and $|E_z|$, respectively, reveals a strong enhancement of the latter (whose maximum is about $4.5$ as opposed to the maximum of the former which is about $0.5$) and this is an essential further signature of the plasmonic resonance. For completeness we have also plotted the amplitude $|1+E_R|$ which corresponds to the superposition of the generated vortex with an additional right hand circularly polarized plane wave. The vortical nature of $E_R$ (i.e. its topological charge is two) is strikingly evident from the spiral shape of the resulting interference which, due to the plasmon resonance, is also radially modulated thus providing an highly nontrivial pattern.

\begin{figure*}
\center
\includegraphics[width=0.95\textwidth]{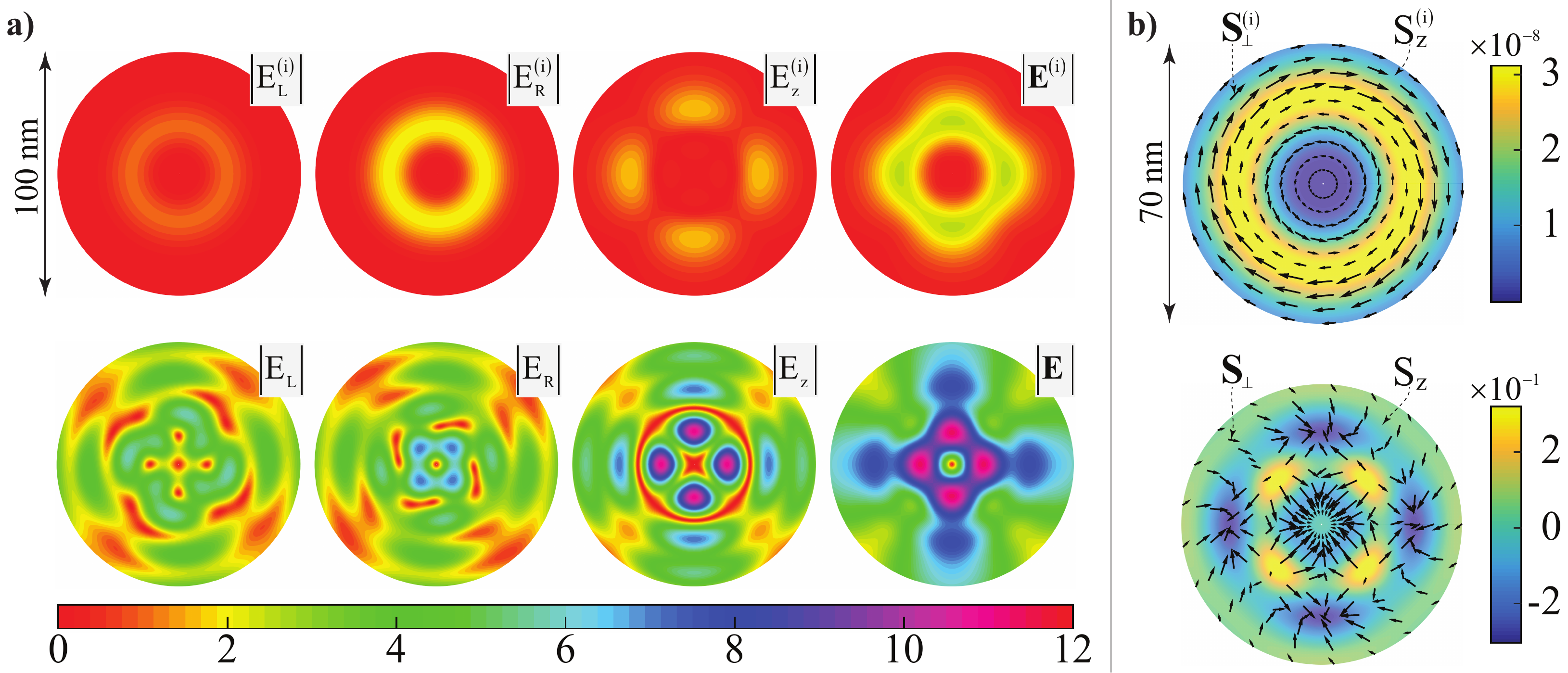}
\caption{Vortex mixing at the wavelength $\lambda = 1.55 \: \mu$m from a graphene sheet whose Fermi energy is $1.14$ eV. (a) Amplitudes of the electric field components and magnitudes of the incoming (upper row) and scattered (lower row) field. The richness of subwavelength spatial details of the scattered patterns is due the superposition of two vortices with different topological charges for each component and to the plasmon resonance accompanying the generation of SPPs. (b) Pictorial representation of the transverse part (as a vector field) and of the longitudinal component (as a color plot) of the incoming and scattered Poynting vectors.}
\end{figure*}

In Fig.3 we again focus on the above considered field with $w=9 \cdot 10^{-3} \: \lambda = 14$ nm and we provide a comparison between the energy flows of the incoming and scattered field. From the expression of the magnetic field derived in Appendix \ref{APP-field}, it turns out that in these conditions both the incoming and scattered magnetic fields have the same vortical structure of the electric field in Eq.(\ref{scatVort}) and this implies that the time-averaged Poynting vector has circularly symmetric cylindrical components, i.e ${\bf{S}}\left({\bf r}_\bot\right) = S_r \left(r_\bot\right) {\bf{e}}_r  + S_\varphi \left(r_\bot\right) {\bf{e}}_\varphi   + S_z \left(r_\bot\right){\bf{e}}_z$. In Figs.3a we plot the cylindrical components of the incoming Poynting vector and we note that $S^{(i)}_\varphi$ is much greater than $S^{(i)}_r$ and $S^{(i)}_z$ so that ${\bf S}^{(i)}$ is essentially an azimuthal field. In Fig.3b we consider a disk centered at ${\bf r} = {\bf 0}$ and diameter of $56$ nm, depicting the transverse part ${\bf{S}}_\bot^{(i)} = S_r^{(i)}  {\bf{e}}_r  + S_\varphi^{(i)} {\bf{e}}_\varphi$ as a vector field and the longitudinal component $S_z^{(i)}$ as a color plot. This pictorially shows that the energy flow of the incoming field mainly circulates around the $z$ axis, a property which is consistent with the circular symmetry and the left hand circular polarization of such field. In Fig.3c we plot the cylindrical components of the scattered Poynting vector and a comparison with Fig.3a reveals a particularly evident redistribution of the energy flow accompanying the vortex generation process. The most striking features are that all the three components are much higher than those of the incoming field, their profile shows a ripple and their amplitudes are mutually comparable. Such features are entailed by the interplay between graphene SOI of light and plasmon resonance. It is well known that the excitation of SPPs is accompanied by a non-trivial energy flow occurring in the plane containing the plasmon wavevector and the normal unit vector. In our case the SPP has a radial distribution and accordingly it affects the $S_r$ and $S_z$ components of the scattered field. The fact that there is a considerable enhancement of such components of the Poynting vector does not evidently violate power conservation of the scattering process since the SPP dramatically affects only the evanescent portion of the field which does not contribute to the overall power carried by the field. In Fig.3d, in analogy with Fig.3b, we depict ${\bf{S}} = S_r  {\bf{e}}_r  + S_\varphi {\bf{e}}_\varphi$ and $S_z$ pertaining the scattered field to pictorially show the energy flow redistribution accompanying vortex generation. The occurrence of a prominent longitudinal component $S_z$ in the scattered field entails a remarkable feature of the transverse part ${\bf{S}}_\bot$: its stream lines connecting the rings where $S_z$ is maximum to the ones where $S_z$ is minimum owing to the divergence-free property of the Poynting vector in vacuum, i.e. $\nabla \cdot {\bf S} = 0$.

\subsection{Vortex Mixing}

As a second application of graphene SOI of light, we discuss the mixing of optical vortices with ensuing generation of complex deep subwavelength optical lattices. Consider an impinging monochromatic field of wavelength $\lambda = 1.55 \: \mu$m with spatial profile at the graphene sample (with Fermi energy $E_{\rm F} = 1.14$ eV)
\begin{eqnarray} \label{incVort}
 {\bf{E}}_ \bot^{(i)}  \left( {{\bf{r}}_ \bot  ,0} \right) &=& E_L^{(i)} \left( {{\bf{r}}_ \bot  } \right){\bf{e}}_L  + E_R^{(i)} \left( {{\bf{r}}_ \bot  } \right){\bf{e}}_R  =  \nonumber \\
  &=& E_0 \frac{{e^2 }}{4}\left( {\frac{{r_ \bot  }}{w}} \right)^4 e^{ - \left( {\frac{{r_ \bot  }}{w}} \right)^2 } \left( {e^{i\varphi } {\bf e}_L  + 2e^{ - i\varphi } {\bf e}_R } \right). \nonumber \\
\end{eqnarray}
This field is very much different from the one considered in the vortex generation process of Sec.\ref{VG} in that it has both left and right hand circularly polarized components and, besides, such components carry the topological charges $+1$ and $-1$, respectively. The fields $E_L^{(i)} {\bf{e}}_L $ and $E_R^{(i)}{\bf{e}}_R$ are genuinely vortex fields with well defined values of orbital and spin angular momenta since they are the superposition of only the basis vortices $(m,s)=(+1,+1)$ (with total angular momentum $j=+2$) and $(m,s)=(-1,-1)$ (with total angular momentum $j=-2$), respectively. As a consequence the incoming field ${\bf{E}}_ \bot^{(i)}$  has not well defined values of the angular momenta but rather it is the superposition of two vortex fields. The spatial profile of $E_L^{\left( i \right)}$ is the same as the one considered in Sec.\ref{VG} whereas $E_R^{\left( i \right)}$ has an amplitude which is twice the other (again $E_0$ is irrelevant and we hereafter set it equal to one).

Following the procedure described in Appendix \ref{APP-field}, the components of the field scattered by the graphene sheet is
\begin{eqnarray} \label{mixVort}
 E_L \left( {{\bf{r}}_ \bot  } \right) &=& E_{LL} \left( r_\bot \right) e^{i\varphi }  + E_{LR} \left( r_\bot  \right)e^{ - i3\varphi }, \nonumber \\
 E_R \left( {{\bf{r}}_ \bot  } \right) &=& E_{RL} \left( r_\bot  \right)e^{i3\varphi } + E_{RR} \left( r_\bot  \right)e^{ - i\varphi }, \nonumber \\
 E_z \left( {{\bf{r}}_ \bot  } \right) &=& E_{zL} \left( r_\bot  \right)e^{i2\varphi } + E_{zR} \left( r_\bot  \right)e^{ - i2\varphi },
\end{eqnarray}
whose amplitudes have close form integral expressions which can be numerically evaluated. Note that the left and right hand circularly polarized components display additional vortex contributions of topological charges $-3$ and $+3$, respectively. These novel contributions are produced by the vortex transitions
\begin{eqnarray}
(m,s) &=& (+1,+1) \rightarrow (+3,-1), \nonumber \\
(m,s) &=& (-1,-1) \rightarrow (-3,+1)
\end{eqnarray}
triggered by graphene SOI of light (each separately conserving the total angular momenta $j=+2$ and $j=-2$). Therefore the scattering produces a mixing of the incoming vortices, each circular component being the superposition of different vortices. Note that the longitudinal component is the superposition of two terms with topological charges $+2$ and $-2$ so that, again, if we associate the spin value $s=0$ to $E_z$, these two contributions have total angular momenta $j=+2$ and $j=-2$, respectively.

In Fig.4 we consider the mixing of the vortices of Eq.(\ref{incVort}) produced by graphene SOI of light for $w=9 \cdot 10^{-3} \: \lambda = 14$ nm. In Fig.4a we plot all the relevant amplitudes on a disk centered at ${\bf r}_\bot = {\bf 0}$ and a diameter of $100$ nm using for all of them the same color scale reported on the bottom for comparison purposes. The first and second rows of Fig.4a are related to the incoming and scattered field, respectively. The amplitudes $\left| {E_L^{\left( i \right)} } \right|$ and $\left| {E_R^{\left( i \right)} } \right|$ have circularly symmetric profiles, the latter being twice the former, whereas $\left| {E_z^{\left( i \right)} } \right|$ has a four lobe structured profile whose magnitude is comparable with the transverse components due to the tight transverse confinement of the field. The resulting electric field magnitude $|{\bf{E}}^{(i)}| = \sqrt{|E_L^{(i)}|^2  + |E_R^{(i)}|^2  + |E_z^{(i)}|^2}$ has an overall ring structure which is not circularly symmetric. The above discussed vortex mixing dramatically affects the amplitudes $\left| {E_L } \right|$ and $\left| {E_R } \right|$ since their patterns are very much different from the corresponding incoming ones and are equipped with very deep sub-wavelength spatial features. Even more complex profiles are observed in the amplitudes $|E_z|$ and $|{\bf E}| = \sqrt{|E_L|^2 + |E_R|^2 + |E_z|^2 }$. The richness of spatial details of such patterns ensues from two basic ingredients. First, each component is the superposition of two vortices with different topological charges so that their interference entails an involved angular structuring. Note that the two vortices in each component have  topological charges differing by $\pm 4$ so that the interference pattern is a periodic function of $4\varphi$ thus explaining the fact that all the reported patterns are left invariant by a rotation of $\pi/2$. Second, due to the tight transverse confinement of the field, in analogy to the vortex generation discussed in Sec.\ref{VG}, vortex mixing is here accompanied by the excitation of SPPs with distinct radial modulation patterns illustrated in the lower row of Fig.4a. In addition plasmon resonance also yields a remarkable enhancement of the longitudinal component $|E_z|$ whose lobes reach a maximum five times larger than the maximum of $\left| {{\bf{E}}^{\left( i \right)} } \right|$. All these features merge to yield a very structured profile of $\left| {\bf{E}} \right|$ whose square has lobes which are about 40 times stronger than the maximum of $\left| {\bf{E}}^{(i)} \right|$.

\begin{figure*}
\center
\includegraphics[width=0.95\textwidth]{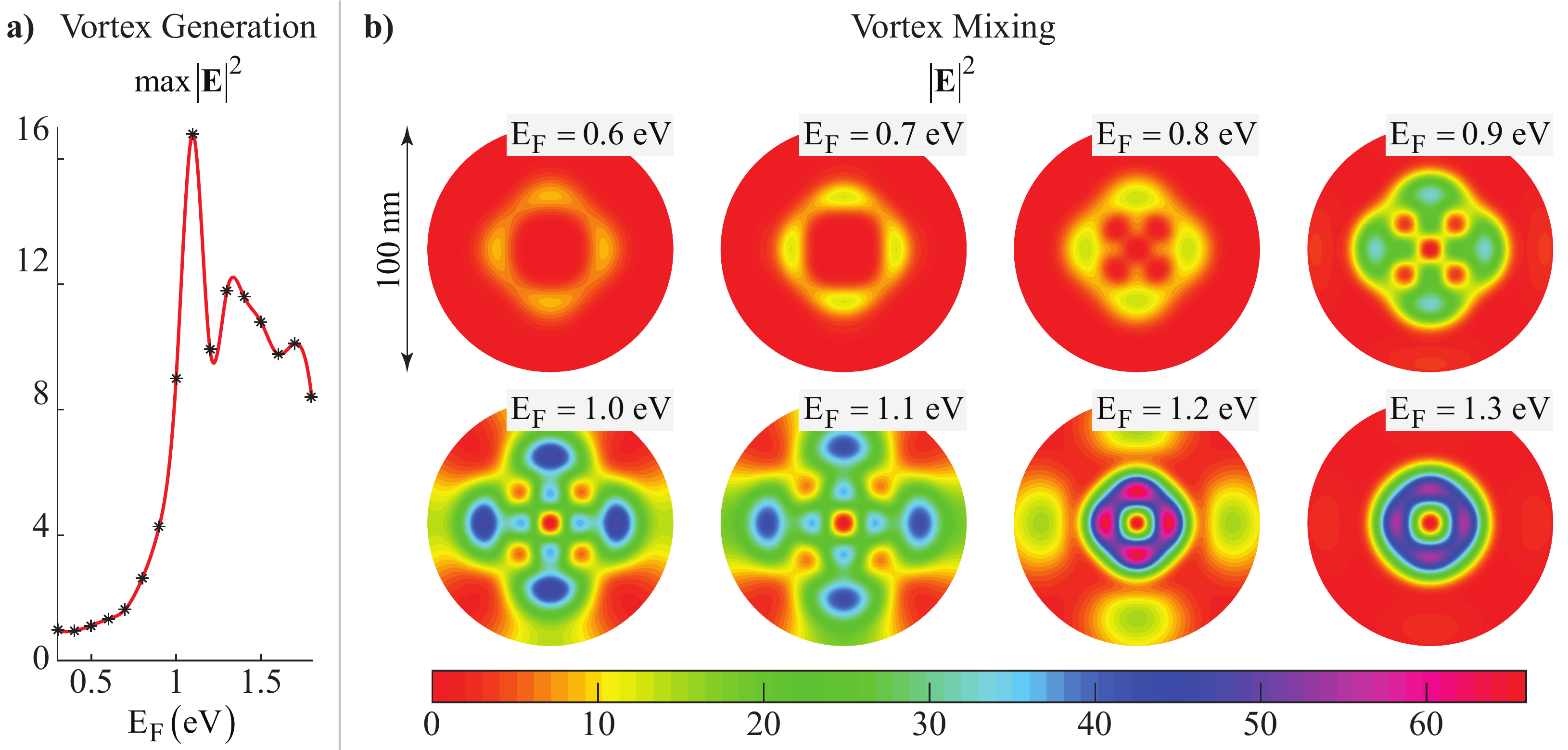}
\caption{Tunability of SOI of light in graphene. (a) Maximum of $|{\bf E}|^2$ in the same vortex generation process examined in Fig. 2c for different Fermi energies. Vortex generation efficiency dramatically raises for $E_{\rm F} > 1$ where the plasmonic resonance spectral profile overlaps the incoming field spectrum. (b) Spatial profile of $|{\bf E}|^2$ in the same vortex mixing process considered in Fig.4a for different values of the Fermi energy $E_{\rm F}$. In this process, external control enables the tuning of both the amplitude and the spatial shape of the scattered field through the Fermi energy. In all figures above the excitation amplitude is set to unity.}
\end{figure*}

Such a remarkable qualitative field transformation operated by the atomically thin graphene sheet is accompanied by an equally strong spatial redistribution of the energy flow. In Fig.4b we compare the Poynting vectors of the incoming and transmitted fields by illustrating their transverse parts ${\bf{S}}_\bot^{(i)}$ and ${\bf{S}}_\bot$ as vector fields and their longitudinal components $S_z^{(i)}$ and $S_z$ as color plots on a disk centered at ${\bf r} = {\bf 0}$ and diameter of $70$ nm. The energy flow of the incoming field mainly circulates around the $z$ axis since $S_z^{(i)}$ is negligible, a feature which ensues from the vortical nature of the circular components of such a field. On the contrary, in the scattered field both ${\bf{S}}_\bot$ and $S_z$ exhibit structured profiles very much different from the incoming ones. Note that $S_z$ is very much stronger than $S_z^{(i)}$ and it exhibits, in addition to an overall ring structure, four lobes surrounding the central dark spot. Again, as in the above considered vortex generation process, the strong longitudinal component $S_z$ of the scattered field arise from the divergence-free nature of the Poynting vector in vacuum (i.e. $\nabla \cdot {\bf S} = 0$), implying that the stream lines of ${\bf S}_\bot$ connect regions where $S_z$ is maximum to regions where $S_z$ is minimum.

\section{Discussion}

The SOI of light discussed above, including vortex generation and mixing, is the main physical process enabling the simultaneous entanglement of spin and orbital angular momentum components of photons. Currently, state-of-the-art macroscopic {\it q -}plates \cite{MarrucciJOPT2011} are being extensively used to encode information into hyper-entangled {\it qdits} of high dimension for quantum cryptography \cite{Bechmann2000,Cerf2002,Mirho2015}. While such game-changing devices can even be miniaturized to some extent, their size is inherently limited by the optical wavelength hampering functionality at the nanoscale, which would be desirable for the development of quantum computers. Recently, gold-based metasurfaces of subwavelength thickness have been shown to overcome this limitation thanks to the plasmonic confinement of the metal and are considered as a viable platform for quantum computation with {\it qdits} \cite{Karimi2014}. However, such devices require advanced fabrication techniques for the realization of involved and precise nanostructured arrays and inherently lack active tunability.

In view of such current limitations and our findings, graphene has all the features for aspiring to become the best material for quantum computation at the nanoscale. Indeed, as discussed in Sec. IIIB and illustrated in Fig. 2, such an atomically thin medium is able to generate efficiently vortices thanks to plasmon-enhanced SOI of light. A striking feature of such a generation process is that, in addition to the vortex angular modulation, the radial oscillations arising from graphene SPP excitation occur at the deep subwavelength nanometer scale, which is unachievable with {\it q -}plates and metasurfaces.

Such a feature becomes even more relevant in the plasmon-assisted vortex mixing discussed in Sec. IIIC and illustrated in Fig. 4, which enables the generation of deep subwavelength optical lattices. Remarkably, by impinging through the nanotip vortices with distinct topological charge and polarizations, plasmon-assisted SOI of light in graphene enables the engineering of arbitrary lattices with angular and radial features of few nanometers, which is inherently unachievable with standard optical tweezers. In turn, thanks also to the extraordinary field enhancement provided by graphene SPPs, such lattices are promising for devising artificial media at will by pinning cold atoms in the desired pattern. Indeed, the time-averaged optical force operated by the electromagnetic field on a generic atom or molecule with polarizability $\alpha>0$ is given by ${\bf F} = (1/2)\alpha\nabla|{\bf E}|^2$ and thus, e.g., the optical lattice depicted in Fig. 4a would arrange atoms/molecules into a square array resulting from the peaks of $|{\bf E}|^2$.

In this context, a further feature that is unique to graphene as compared to other photonic materials lies in the external tunability through the injection of charge carriers, which enables the active manipulation of vortex generation efficiency and subwavelength optical lattices through an external gate voltage. Tunability mainly arises from the fact that the spectral position and width of the plasmonic resonance are highly dependent on the Fermi energy of the graphene sample (see Fig. 1d), and in Fig. 5 we illustrate the potential offered by external control. In Fig. 5a we consider the same vortex generation process examined in Fig. 2c and we plot the maximum of $|{\bf E}|^2$ for Fermi energies $E_{\rm F}$ in the range $0.3 \div 1.8$ eV. For small values of $E_{\rm F}$ ($< 0.8$ eV) vortex generation is not very efficient with ${\rm max} \: |{\bf E}|^2$ close to $1$ since the plasmonic resonance is located at $k_\bot > 100 \: k_0$, far outside the main spectral content of the incoming field (see the right inset of Fig. 2a). For higher Fermi energies, the plasmonic resonance efficiently enhances vortex generation since its spectral profile fully overlaps with the incoming field spectrum. From an application point of view, such behavior could be extremely useful for achieving ultrafast amplitude modulation of deep-subwavelength confined vortices. Note that the decrease of ${\rm max} \: |{\bf E}|^2$ for $E_{\rm F} > 1.1$ eV is not due to some kind of plasmonic resonance quenching, but it is rather associated to the small lateral spectral lobe of ${\tilde E}_L^{(i)}$ (see again the right inset of Fig. 2a). In Fig.5b we examine the spatial profile of $|{\bf E}|^2$ of the same vortex mixing process considered in Fig.4a for different values of the Fermi energy $E_{\rm F}$. In this process plasmonic tunability exhibits an even more dramatic phenomenology since $|{\bf E}|^2$ displays both different amplidude and different shapes at different Fermi energies. In turn, the atomically-thin graphene sheet may constitute the core of a device able to generate extreme deep-subwavelength optical lattices with shape tunable in real time by means of the external bias voltage.

\begin{figure}
\center
\includegraphics[width=0.45\textwidth]{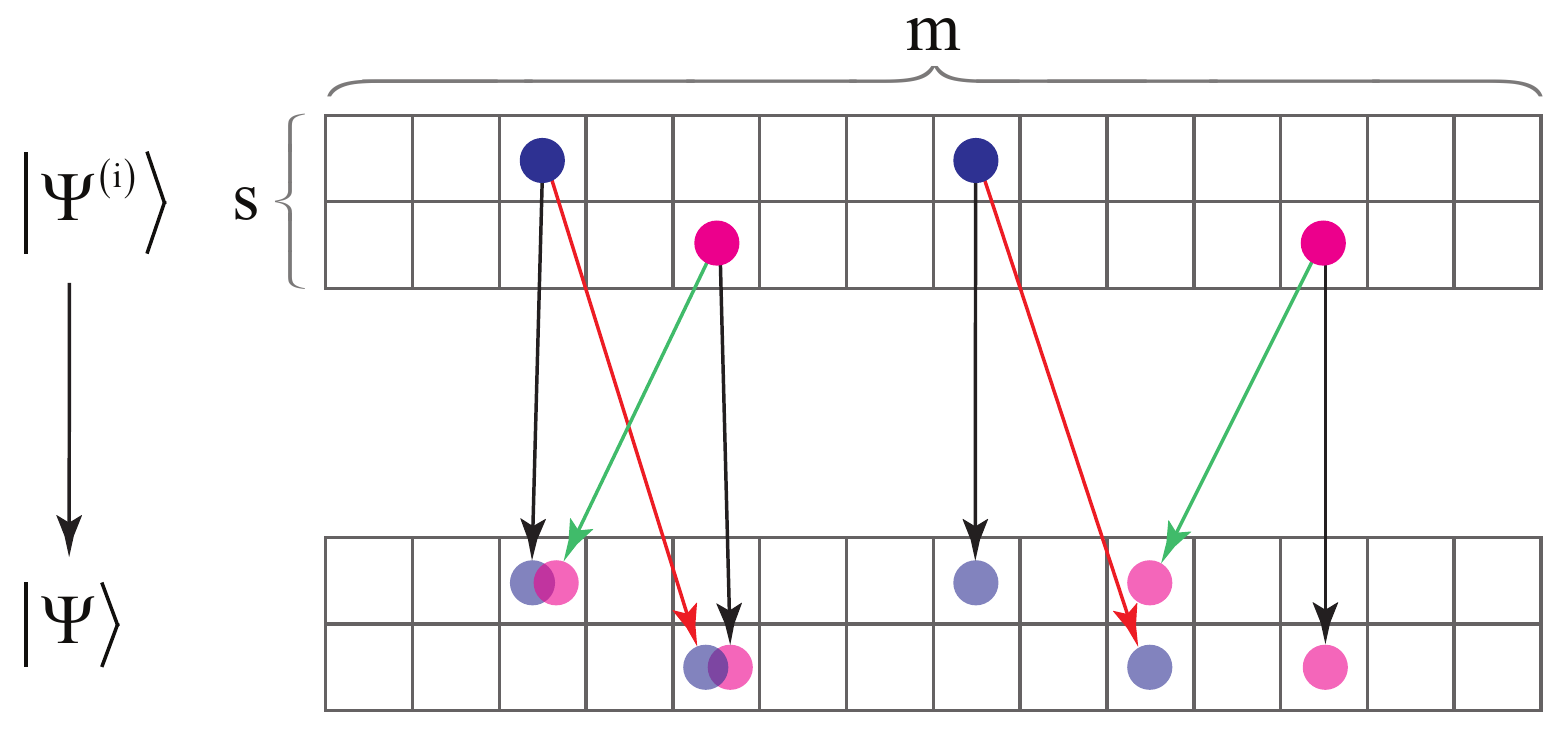}
\caption{Pictorial illustration of information processing based on the SOI of light. Information can be encoded in the radiation state through its vortex $(m,s)$ distribution, which electrically tunable change upon scattering actively manipulates the information stored.
}
\end{figure}

Although our formalism and predictions are fully classical, they can be straightforwardly generalized to the quantum regime since all the operators introduced to describe light-matter interaction are either diagonal in the eigenstates of the total angular momentum or are expressed in terms of ladder operators (see Sec. IIIA and Appendices). In turn, in view of the results discussed above, the extended doped graphene sheet considered in our calculations can perform tunable logic operations with single photons carrying information in {\it qdits}. For example, in Fig.6 we pictorially illustrate a possible way a exploiting the SOI of light to achieve nontrivial information manipulation. The information can be encoded in the radiation state through its vortex $(m,s)$ distribution. The change of the vortex distribution through scattering effectively amounts, using the Shannon terminology, to the conversion of latent information (stored in $\left| {\Psi ^{\left( i \right)} } \right\rangle$) into manifest information (stored in $\left| \Psi  \right\rangle$), i.e. information processing occurs. Such manipulation functionality encompasses various different channels since different bits can be activated by vortex generation or switched off by suitable interference entailed by vortex mixing. In addition, the range of possible channels is also increased by tunability, thus enabling to control efficiency and features of the overall information processing functionality.

\section{Conclusions}

In summary, without taking any approximation, we develop a new theoretical framework accounting for the spin orbit interaction of arbitrary light fields, showing that their tight confinement along with the plasmonic resonance supported by doped extended graphene leads to the generation of optical vortices at the nanoscale with unprecedentedly high efficiency and deep-subwavelength spatial features. Remarkably, in spite of the atomic thickness of such a two-dimensional material, we demonstrate that it outperforms {\it q -}plates and metasurfaces with the further advantage of nano-operation and electrical tunability. While the novel theoretical model derived captures the spin orbit interaction of light mediated by an arbitrary 2D material, we show that extended graphene is ideal since it hosts surface plasmon polaritons with long lifetime and high quality factor, increasing the vortex generation efficiency. Furthermore, thanks to the mixing and interference of distinct vortices, we demonstrate the ability of graphene to generate deep subwavelength optical lattices of arbitrary shape and pace of few nanometers, enabling to devise artificial media at will. Although future work is required to extend our results to the quantum regime, we envisage that they will constitute a solid theoretical ground for the development of nano-scaled active elements and logic gates for enhanced quantum computation based on hyper-entangled photon states.

\section{Acknowledgments}

A.C. and C.R. acknowledge support from U.S. Army International Technology Center Atlantic for financial support (Grant No. W911NF-14-1-0315). A.M. acknowledges support from the Rita Levi Montalcini Fellowship (Project No. PGR15PCCQ5) funded by the Italian Ministry of Education, Universities and Research (MIUR).

\appendix
\section{Eigenvectors of $\hat K$, $\hat L$ and $\hat S$: the Vortex Basis} \label{APP-eigenvectors}
Let us consider the simultaneous eigenvalue problem of the operators $\hat K$, $\hat L$ and $\hat S$
\begin{eqnarray} \label{sim-eig}
\begin{array}{l}
 \hat K\left| {k_ \bot  ,m,s} \right\rangle  = u\left( {k_ \bot  } \right)\left| {k_ \bot  ,m,s} \right\rangle,  \\
 \hat L\left| {k_ \bot  ,m,s} \right\rangle  = m\left| {k_ \bot  ,m,s} \right\rangle,  \\
 \hat S\left| {k_ \bot  ,m,s} \right\rangle  = s\left| {k_ \bot  ,m,s} \right\rangle.  \\
 \end{array}
\end{eqnarray}
Since the operators $\hat K$ and $\hat L$ have no effect on the spin degrees of freedoms and the spin operator $ \hat S$ has no effect on the orbital state, it is possible to set
\begin{equation} \label{eigenvect}
\left| {k_ \bot  ,m,s} \right\rangle  = \left| {k_ \bot,m} \right\rangle  \otimes \left| s \right\rangle.
\end{equation}
The diagonalization of the spin operator $\hat S_{\rm spi}   = \frac{1}{i}\left( {\left| {{\bf{e}}_x } \right\rangle \left\langle {{\bf{e}}_y } \right| - \left| {{\bf{e}}_y } \right\rangle \left\langle {{\bf{e}}_x } \right|} \right)$ is straightforward and its eigenvalues are $s=+1$ and $s=-1$ with corresponding eigenvectors
\begin{equation} \label{spin-states}
\left| s \right\rangle  = \frac{1}{{\sqrt 2 }}\left( {\left| {{\bf{e}}_x } \right\rangle  + is\left| {{\bf{e}}_y } \right\rangle } \right),
\end{equation}
which are an orthonormal basis of $\mathcal{H}_{\rm spi}$, i.e.
\begin{eqnarray} \label{bas-spi}
 \left\langle {s}
 \mathrel{\left | {\vphantom {s {s'}}}
 \right. \kern-\nulldelimiterspace}
 {{s'}} \right\rangle  &=& \delta _{ss'}, \nonumber  \\
 \sum\limits_{s = -1,1} {\left| s \right\rangle \left\langle s \right|}  &=& \hat I_{\rm spi}.
\end{eqnarray}
In order to find the orbital eigenvectors $\left| {k_\bot,m} \right\rangle$ it is convenient to cast the operator $\hat K$ in a different form. Setting $\hat K = \hat K_{\rm orb}  \otimes \hat I_{\rm spi}$ and using the Weyl representation of the spherical wave \cite{App_Mandle}
\begin{equation}
h\left( {\bf r} \right)= \frac{i}{{2\pi }}\int {d^2 {\bf{k}}_ \bot  } e^{i{\bf{k}}_ \bot   \cdot {\bf{r}}_ \bot  } \frac{{e^{i\sqrt {k_0^2  - k_ \bot ^2 } \left| z \right|} }}{{\sqrt {k_0^2  - k_ \bot ^2 } }},
\end{equation}
we have
\begin{eqnarray} \label{K-orb}
 \hat K_{\rm orb}  &=&  - \frac{{k_0 }}{{4\pi ^2 }}\int {d^2 {\bf{r}}_ \bot  } \int {d^2 {\bf{r}}'_ \bot  }  \times  \nonumber \\
  &\times& \int {d^2 {\bf{k}}_ \bot  } \frac{{e^{i{\bf{k}}_ \bot   \cdot \left( {{\bf{r}}_ \bot   - {\bf{r}}'_ \bot  } \right)} }}{{\sqrt {k_0^2  - k_ \bot ^2 } }}\left| {{\bf{r}}_ \bot  } \right\rangle \left\langle {{\bf{r}}'_ \bot  } \right| = \nonumber \\
   &=&  \int {d^2 {\bf{k}}_ \bot  } \frac{ - k_0}{{\sqrt {k_0^2  - k_ \bot ^2 } }}\left| {{\bf{k}}_ \bot  } \right\rangle \left\langle {{\bf{k}}_ \bot  } \right| = \nonumber \\
  &=&  - \left( {1 + \frac{1}{{k_0^2 }}\hat D_ \bot ^2 } \right)^{ - 1/2},
\end{eqnarray}
where $\left| {{\bf{k}}_ \bot  } \right\rangle  = \frac{1}{{2\pi }}\int {d^2 {\bf{r}}_ \bot  } e^{i{\bf{k}}_ \bot   \cdot {\bf{r}}_ \bot  } \left| {{\bf{r}}_ \bot  } \right\rangle$ are the plane waves states (which form an orthonormal basis of $\mathcal{H}_{\rm orb}$, i.e. $\left\langle {{{\bf{k}}_ \bot  }} \mathrel{\left | {\vphantom {{{\bf{k}}_ \bot  } {{\bf{k}}'_ \bot  }}}  \right. } {{{\bf{k}}'_ \bot  }} \right\rangle  = \delta \left( {{\bf{k}}_ \bot   - {\bf{k}}'_ \bot  } \right)$, $\int {d^2 {\bf{k}}_ \bot  } \left| {{\bf{k}}_ \bot  } \right\rangle \left\langle {{\bf{k}}_ \bot  } \right| = I_{\rm orb}$)),
\begin{equation}
\hat D_ \bot ^2  = \hat D_x^2  + \hat D_y^2
\end{equation}
is the transverse Laplacian operator whose eigenvectors are the plane wave states, i.e.
\begin{equation}
\hat D_ \bot ^2 \left| {{\bf{k}}_ \bot  } \right\rangle  =  - k_ \bot ^2 \left| {{\bf{k}}_ \bot  } \right\rangle.
\end{equation}
Equation (\ref{K-orb}) reveals that $\hat K_{\rm orb}$ is strictly a function of the transverse Laplacian operator and therefore $\left| {k_\bot,m} \right\rangle$ are the simultaneous eigenvectors of $\hat D_ \bot ^2$ and $\hat L_{\rm orb} = \frac{1}{i}\left( {\hat X\hat D_y  - \hat Y\hat D_x } \right)$, i.e.
\begin{eqnarray}
 \hat D_ \bot ^2 \left| {k_ \bot  ,m} \right\rangle  &=&  - k_ \bot ^2 \left| {k_ \bot  ,m} \right\rangle, \nonumber  \\
 \hat L_{\rm orb} \left| {k_ \bot  ,m} \right\rangle  &=& m\left| {k_ \bot  ,m} \right\rangle.
\end{eqnarray}
Expanding the eigenvector on the basis $\left| {{\bf{r}}_ \bot  } \right\rangle$ by setting $\left| {k_ \bot  ,m} \right\rangle  = \int {d^2 {\bf{r}}_ \bot  } f_{k_ \bot  ,m} \left( {{\bf{r}}_ \bot  } \right)\left| {{\bf{r}}_ \bot  } \right\rangle$, we get
\begin{eqnarray}
 \nabla _ \bot ^2 f_{k_ \bot  ,m}  &=&  - k_ \bot ^2 f_{k_ \bot  ,m}, \nonumber \\
 \frac{\partial f_{k_ \bot  ,m}}{{\partial \varphi }}  &=& imf_{k_ \bot  ,m}.
\end{eqnarray}
where ${\bf{r}}_ \bot   = r_ \bot  \left( {\cos \varphi {\bf{e}}_x  + \sin \varphi {\bf{e}}_y } \right)$ so that the eigenfunctions $f_{k_ \bot  ,m}$ are the well-known cylindrical harmonics \cite{Schwinger} where $k_\bot$ is any positive real number and $m$ is any integer number. Accordingly the eigenvectors are
\begin{equation}
\left| {k_ \bot  ,m} \right\rangle  = \int {d^2 {\bf{r}}_ \bot  } \sqrt {\frac{{k_ \bot  }}{{2\pi }}} J_m \left( {k_ \bot  r_ \bot  } \right)e^{im\varphi } \left| {{\bf{r}}_ \bot  } \right\rangle,
\end{equation}
where $J_m (\zeta)$ is the Bessel function of the first kind of order $m$ and they form an orthonormal basis of $\mathcal{H}_{\rm orb}$, i.e.
\begin{eqnarray} \label{bas-orb}
 \left\langle {{k_ \bot  ,m}}
 \mathrel{\left | {\vphantom {{k_ \bot  ,m} {k'_ \bot  ,m'}}}
 \right. \kern-\nulldelimiterspace}
 {{k'_ \bot  ,m'}} \right\rangle  = \delta \left( {k_ \bot   - k'_ \bot  } \right) \delta _{mm'}, \nonumber  \\
 \int\limits_0^\infty  {dk_ \bot  } \sum\limits_{m =  - \infty }^\infty  {\left| {k_ \bot  ,m} \right\rangle } \left\langle {k_ \bot  ,m} \right| = I_{\rm orb}.
\end{eqnarray}
We conclude that the eigenvectors of Eq.(\ref{eigenvect}) are
\begin{eqnarray}
\left| {k_ \bot  ,m,s} \right\rangle  &=& \int {d^2 {\bf{r}}_ \bot  } \sqrt {\frac{{k_ \bot  }}{{2\pi }}} J_m \left( {k_ \bot  r_ \bot  } \right)e^{im\varphi } \left| {{\bf{r}}_ \bot  } \right\rangle \otimes \nonumber \\
  &\otimes& \frac{1}{{\sqrt 2 }}\left( {\left| {{\bf{e}}_x } \right\rangle  + is\left| {{\bf{e}}_y } \right\rangle } \right)
\end{eqnarray}
and, due to Eqs.(\ref{bas-spi}) and (\ref{bas-orb}), they form an orthonormal basis of $\mathcal{H}$, i.e.
\begin{eqnarray}
 \left\langle {k_ \bot  ,m,s} \right.|\left. {k'_ \bot  ,m',s'} \right\rangle  &=& \delta \left( {k_ \bot   - k'_ \bot  } \right)\delta _{mm'} \delta _{ss'}, \nonumber  \\
 \sum\limits_{k_ \bot  ,m,s} {|\left. {k_ \bot  ,m,s} \right\rangle \left\langle {k_ \bot  ,m,s} \right.|\;}  &=& \hat I,
\end{eqnarray}
where, for notation convenience, we hereafter set
\begin{equation} \label{shorthand}
\sum\limits_{k_ \bot  ,m,s} {}  \equiv \int\limits_0^{ + \infty } {dk_\bot} \sum\limits_{m =  - \infty }^\infty  {\sum\limits_{s = - 1,1} }.
\end{equation}
In the representation induced by the eigenvectors $\left| {k_ \bot  ,m,s} \right\rangle$, the operators $\hat K$, $\hat L$ and $\hat S$ are
\begin{eqnarray} \label{KLS}
 \hat K &=& \sum\limits_{k_ \bot  ,m,s} {} u \left(k_\bot\right)
  |\left. {k_ \bot  ,m,s} \right\rangle \left\langle {k_ \bot  ,m,s} \right.|, \nonumber  \\
 \hat L &=& \sum\limits_{k_ \bot  ,m,s} {m|\left. {k_ \bot  ,m,s} \right\rangle \left\langle {k_ \bot  ,m,s} \right.|}, \nonumber  \\
 \hat S &=& \sum\limits_{k_ \bot  ,m,s} {s|\left. {k_ \bot  ,m,s} \right\rangle \left\langle {k_ \bot  ,m,s} \right.|},
\end{eqnarray}
where
\begin{equation}
u\left( {k_ \bot  } \right) =  - \left( {1 - \frac{k_ \bot ^2}{ k_0^2} } \right)^{ - 1/2}.
\end{equation}

A deeper understanding of the vortex representation can be gained from some geometrical consideration. In the presence of graphene, only rotations around the $z$ axis are allowed and hence the symmetry group is $SO(2)$. This group has infinitely many irreducible representations which are all one-dimensional since it is Abelian. First, the unitary operators $e^{ - i \hat L \vartheta }$ provide a representation of $SO(2)$ on $\mathcal{H}$ and, since $e^{ - i\hat L\vartheta } \left| {k_ \bot  ,m,s} \right\rangle  = e^{ - im\vartheta } \left| {k_ \bot  ,m,s} \right\rangle$, each vector $\left| {k_ \bot  ,m,s} \right \rangle$ spans a one-dimensional invariant subspace, i.e. it is the basis of the one-dimensional irreducible representation whose character is $e^{ - im\vartheta }$. This observation explains why the orbital angular momentum $m$ is not bounded as opposed to the constraint $-l<m<l$ pertaining the general $(2l+1)-$dimensional irreducible representation of the rotation group $SO(3)$. Second, the operators $e^{ - i\hat S\vartheta }$ provide a different representation of $SO(2)$ on $\mathcal{H}$ and, since $e^{ - i\hat S\vartheta } \left| {k_ \bot  ,m,s} \right\rangle  = e^{ - is\vartheta } \left| {k_ \bot  ,m,s} \right\rangle$, the vectors $\left| {k_ \bot  ,m, + 1} \right\rangle$ and $\left| {k_ \bot  ,m, - 1} \right\rangle$ are bases of the two irreducible representations whose characters are $e^{ - i\vartheta }$ and $e^{i\vartheta }$, respectively. This situation should be compared with that of a spin-1 particle whose spin state is three-dimensional and whose spin operator $\hat S_z$ has three eigenvectors providing the three irreducible representations of $SO(2)$ whose characters are $e^{ - i\vartheta }$, $e^{i\vartheta }$ and $1$. As a matter of fact such three eigenvectors correspond in the 3D cartesian space to the unit vectors ${\bf{e}}_L$, ${\bf{e}}_R$ and ${\bf{e}}_z$. In the presence of graphene, the transverse electric field is not coupled with its longitudinal component $E_z$ (see Eqs.(\ref{scat})) so that it has been possible to deal with a two dimensional polarization (spin) state space and with the spin operator $\hat{S}$ which is the restriction of its three-dimensional counterpart $\hat S_z$ to the transverse space. Third, the unitary rotation operators $e^{ - i \hat J \vartheta }$ on $\mathcal{H}$ provide a further representation of $SO(2)$. Since each subspace $\mathcal{E} \left( {k_ \bot  ,j} \right)$ of Eq.(\ref{E(kj)}) is an eigenspace of $\hat{J}$, it is invariant invariant for the rotation operators, and hence it provides the carrier space of a two-dimensional reducible representation of $SO(2)$. Now $e^{ - i\hat J\vartheta } \left| {k_ \bot  ,j - 1,s + 1} \right\rangle  = e^{ - ij\vartheta } \left| {k_ \bot  ,j - 1,s + 1} \right\rangle$ and $e^{ - i\hat J\vartheta } \left| {k_ \bot  ,j + 1,s - 1} \right\rangle  = e^{ - ij\vartheta } \left| {k_ \bot  ,j + 1,s - 1} \right\rangle$ so that each one of these two-dimensional representation is the direct sum of two one-dimensional irreducible representation having the same character $e^{ - ij\vartheta }$.

\section{Orbital and spin ladder opertors} \label{APP-ladder}
Consider the two operators
\begin{eqnarray} \label{L+L-}
 \hat L_ +  &=& \left( {\hat D_x  + i\hat D_y } \right) \otimes \hat I_{\rm spi}, \nonumber \\
 \hat L_ -  &=& \left( {\hat D_x  - i\hat D_y } \right) \otimes \hat I_{\rm spi}.
\end{eqnarray}
which are straightforwardly seen to satisfy the commutation relations
\begin{eqnarray} \label{alge}
 [ \hat L_ +  ,\hat L_ -  ] &=& 0, \nonumber \\
 { [ \hat L, \hat L_\pm  ]} &=&  \pm \hat L_\pm.
\end{eqnarray}
The second of these equations shows that $\hat L_+$ and $\hat L_-$ are orbital ladder operators, raising and lowering the orbital angular momentum $m$, respectively, by one unit. By letting the operators $\hat L_ \pm$ to act on the basis vectors $\left| {k_ \bot  ,m,s} \right\rangle$, and using polar coordinates ${\bf{r}}_ \bot   = r_ \bot  \left( {\cos \varphi {\bf{e}}_x  + \sin \varphi {\bf{e}}_y } \right)$  in the transverse plane, we have
\begin{widetext}
\begin{eqnarray} \label{orb-ladder}
 \hat L_ \pm  \left| {k_ \bot  ,m,s} \right\rangle  &=& \int\limits_0^{ + \infty } {dr_ \bot  } r_ \bot  \int\limits_0^{2\pi } {d\varphi } \;e^{ \pm i\varphi } \left( {\frac{\partial }{{\partial r_ \bot  }} \pm \frac{i}{{r_ \bot  }}\frac{\partial }{{\partial \varphi }}} \right)\sqrt {\frac{{k_ \bot  }}{{2\pi }}} J_m \left( {k_ \bot  r_ \bot  } \right)e^{im\varphi } \left| {{\bf{r}}_ \bot  } \right\rangle  \otimes \left| s \right\rangle = \nonumber  \\
 &=& \sqrt {\frac{{k_ \bot  }}{{2\pi }}} \int\limits_0^{ + \infty } {dr_ \bot  } r_ \bot  \int\limits_0^{2\pi } {d\varphi } \;\left[ {\frac{{\partial J_m \left( {k_ \bot  r_ \bot  } \right)}}{{\partial r_ \bot  }} \mp \frac{m}{{r_ \bot  }}J_m \left( {k_ \bot  r_ \bot  } \right)} \right]e^{i\left( {m \pm 1} \right)\varphi } \left| {{\bf{r}}_ \bot  } \right\rangle  \otimes \left| s \right\rangle = \nonumber \\
 &=&  \mp k_ \bot  \int\limits_0^{ + \infty } {dr_ \bot  } r_ \bot  \int\limits_0^{2\pi } {d\varphi } \;\sqrt {\frac{{k_ \bot  }}{{2\pi }}} J_{m \pm 1} \left( {k_ \bot  r_ \bot  } \right)e^{i\left( {m \pm 1} \right)\varphi } \left| {{\bf{r}}_ \bot  } \right\rangle  \otimes \left| s \right\rangle  = \nonumber \\
&=&  \mp k_ \bot  \left| {k_ \bot  ,m \pm 1,s} \right\rangle,
\end{eqnarray}
\end{widetext}
where the Bessel function identity $J_m '\left( \zeta  \right) \mp mJ_m \left( \zeta  \right)/\zeta  =  \mp J_{m \pm 1} \left( \zeta  \right)$ has been exploited in the third step.

Note that such ladder operators are different from those commonly used when dealing with the full three-dimensional rotation group (i.e. $\hat L_x \pm \hat L_y$). As discussed in Appendix \ref{APP-eigenvectors}, each vector $\left| {k_ \bot  ,m,s} \right\rangle$ spans the one-dimensional carrier space of an irreducible representation of the Abelian group $SO(2)$ (rotations in the plane). Therefore the operators $\hat L_+$ and $\hat L_-$ connect the carrier spaces of different irreducible representation of $SO(2)$ as opposed to the standard ladder operators whose action is restricted to the carrier space of a fixed $(2l+1)$-dimensional irreducible representation of $SO(3)$.

Consider now the two operators
\begin{eqnarray} \label{S+S-}
 \hat S_ +   &=& \hat I_{\rm orb}  \otimes \left| { + 1} \right\rangle \left\langle { - 1} \right|, \nonumber \\
 \hat S_ -   &=& \hat I_{\rm orb}  \otimes \left| { - 1} \right\rangle \left\langle { + 1} \right|,
\end{eqnarray}
where the $\left| { \pm 1} \right\rangle$ are the eigenvectors of the spin operator $\hat{S}_{\rm spi}$. They are straightforwardly seen to satisfy the commutation relations
\begin{eqnarray} \label{alge2}
 [ \hat S_ +  ,\hat S_ -   ] & =& \hat S, \nonumber \\
 { [ \hat S,\hat S_ \pm ] } &=&  \pm 2\hat S_ \pm,
\end{eqnarray}
the second of which shows that $\hat S_ + $ and $\hat S_ - $ are spin ladder operators, raising and lowering the spin $s$, respectively, by two units. The action of these operators on the basis vectors $|\left. {k_ \bot  ,m, s} \right\rangle$ is given by
\begin{eqnarray}
 \hat S_ +  |\left. {k_ \bot  ,m, + 1} \right\rangle  &=& 0, \nonumber \\
 \hat S_ +  |\left. {k_ \bot  ,m, - 1} \right\rangle  &=& |\left. {k_ \bot  ,m, + 1} \right\rangle, \nonumber  \\
 \hat S_ -  |\left. {k_ \bot  ,m, + 1} \right\rangle  &=& |\left. {k_ \bot  ,m, - 1} \right\rangle, \nonumber  \\
 \hat S_ -  |\left. {k_ \bot  ,m, - 1} \right\rangle  &=& 0,
\end{eqnarray}
which can be summarized as
\begin{equation}
\hat S_ \pm  |\left. {k_ \bot  ,m,s} \right\rangle  = \delta _{s, \mp 1} |\left. {k_ \bot  ,m,s \pm 2} \right\rangle,
\end{equation}
From this relation it is particularly evident that the spin raising and lowering are accompanied by the changes $\Delta s = + 2$ and $\Delta s = -2$, respectively.

\section{The interaction operators} \label{APP-inter}
The orbital ladder operators provide a natural factorization of the transverse Laplacian operator since
\begin{equation} \label{transv-lap}
\hat D_\bot^2 \otimes \hat{I}_{\rm spi} = \hat L_ +  \hat L_ - = \hat L_ -  \hat L_ +
\end{equation}
and accordingly, from Eq.(\ref{orb-ladder}) we have
\begin{equation}
\hat L_ +  \hat L_ -  \left| {k_ \bot  ,m,s} \right\rangle  =  - k_ \bot ^2 \left| {k_ \bot  ,m,s} \right\rangle.
\end{equation}
From Eqs.(\ref{alge}) it easily seen that $[ \hat L_ +  \hat L_ - ,\hat J ] = 0$, which states the well-known rotational invariance of the transverse Laplacian operator. From the first of Eqs.(\ref{KLS}), we obtain
\begin{equation}
\hat K =  - \left( {\hat I + \frac{1}{{k_0^2 }}\hat L_ +  \hat L_ -  } \right)^{ - 1/2},
\end{equation}
from which it is evident that
\begin{equation}
 [ {\hat K,\hat L} ] = [ {\hat K,\hat S} ] = [ {\hat K,\hat J}] =0,
\end{equation}
so that the operator $\hat K$ is rotationally invariant (it conserves the total angular momentum) and it also conserves both the orbital and spin angular momenta.

By inverting Eq.(\ref{spin-states}) we get
\begin{eqnarray}
 \left| {{\bf{e}}_x } \right\rangle  &=& \frac{1}{{\sqrt 2 }}\left( {\left| { + 1} \right\rangle  + \left| { - 1} \right\rangle } \right), \nonumber  \\
 \left| {{\bf{e}}_y } \right\rangle  &=& \frac{1}{{\sqrt 2 i}}\left( {\left| { + 1} \right\rangle  - \left| { - 1} \right\rangle } \right)
\end{eqnarray}
which inserted into the first of Eq.(\ref{MK}), after some algebra, yield
\begin{eqnarray}
\hat M &=& 1 + \frac{1}{{2k_0^2 }}\left[ \left( {\hat D_x^2  + \hat D_y^2 } \right) \otimes I_{\rm spi}  \right. + \nonumber \\
&+&  \left. \left( {\hat D_x  + i\hat D_y } \right)^2 \left| { - 1} \right\rangle \left\langle { + 1} \right| + \left( {\hat D_x  - i\hat D_y } \right)^2 \left| { + 1} \right\rangle \left\langle { - 1} \right| \right]. \nonumber \\
\end{eqnarray}
Using Eqs.(\ref{L+L-}), (\ref{transv-lap}) and (\ref{S+S-}), we obtain
\begin{equation}
\hat M = \hat I + \frac{1}{{2k_0^2 }}\left( {\hat L_ +  \hat L_ -   + \hat L_ + ^2 \hat S_ -   + \hat L_ - ^2 \hat S_ +  } \right).
\end{equation}
By using Eqs.(\ref{alge}) and (\ref{alge2}) we obtain the commutation relations
\begin{eqnarray}
 [ {\hat M,\hat L} ] &=& -  [ {\hat M,\hat S} ] =  \frac{1}{{k_0^2 }}\left( {- \hat L_ + ^2 \hat S_ -   + \hat L_ - ^2 \hat S_ +  } \right), \nonumber \\
 {[ {\hat M,\hat J} ] }&=& 0.
\end{eqnarray}
which shows that the operator $\hat M$ is rotationally invariant (it conserves the total angular momentum) but it does not conserve both the orbital and spin angular momenta.

\section{Solution of the LS equation} \label{APP-LS}
In order to solve the LS equation (Eq.(\ref{lip-sch-abs})), we expand both the incoming and unknown states in the vortex basis, i.e.
\begin{eqnarray} \label{expa}
 \left| {\Psi ^{\left( i \right)} } \right\rangle  &=& \sum\limits_{k_ \bot  ,m,s} {\psi _{k_ \bot  ,m,s}^{\left( i \right)} |\left. {k_ \bot  ,m,s} \right\rangle \;}, \nonumber  \\
 \left| \Psi  \right\rangle  &=& \sum\limits_{k_ \bot  ,m,s} {\psi _{k_ \bot  ,m,s} |\left. {k_ \bot  ,m,s} \right\rangle },
\end{eqnarray}
where $\psi _{k_ \bot  ,m,s}^{\left( i \right)}  = \left\langle {{k_ \bot  ,m,s}}  \mathrel{\left | {\vphantom {{k_ \bot  ,m,s} {\Psi ^{\left( i \right)} }}} \right. \kern-\nulldelimiterspace} {{\Psi ^{\left( i \right)} }} \right\rangle$ and ${\psi _{k_ \bot  ,m,s} }$
are unknown. Substituting Eqs.(\ref{expa}) into Eq.(\ref{lip-sch-abs}) and exploiting the ladder operators properties, after some algebra  we get
\begin{eqnarray} \label{recu}
&&\left[ {\left( {1 - \xi u} \right) + \left( {1 - \xi u^{ - 1} } \right)} \right]\psi _{k_ \bot  ,m,s}  +\nonumber \\
&+& \left[ {\left( {1 - \xi u} \right) - \left( {1 - \xi u^{ - 1} } \right)} \right]\psi _{k_ \bot  ,m + 2s, - s}  = 2\psi _{k_ \bot  ,m,s}^{\left( i \right)}, \nonumber  \\
\end{eqnarray}
where $u(k_\bot)$ is the eigenvalue of the operator $\hat K$, as above. Note that each term in Eq.(\ref{recu}) has the same index $k_\bot$ and the same total angular momentum $j=m+s$, so that we are effectively solving the LS equation in each subspace $\mathcal{E}\left( {k_ \bot  ,j} \right)$ in agreement with the rotationally invariance of the system. Substituting the trial solution
\begin{equation}
\psi _{k_ \bot  ,m,s}  = Q\psi _{k_ \bot  ,m,s}^{\left( i \right)}  + P\psi _{k_ \bot  ,m + 2s, - s}^{\left( i \right)}
\end{equation}
into Eq.(\ref{recu}) we obtain a system of two linear equations for $Q$ and $P$ whose solution is
\begin{eqnarray}
 Q &=& \frac{1}{2}\left( {\frac{1}{{1 - \xi u}} + \frac{1}{{1 - \xi u^{ - 1} }}} \right), \nonumber \\
 P &=& \frac{1}{2}\left( {\frac{1}{{1 - \xi u}} - \frac{1}{{1 - \xi u^{ - 1} }}} \right).
\end{eqnarray}
The solution of the LS equation accordingly is
\begin{eqnarray}
 \left| \Psi  \right\rangle  &=& \sum\limits_{k_ \bot  ,m,s} {\psi _{k_ \bot  ,m,s}^{\left( i \right)} Q\left( {k_ \bot  } \right)|\left. {k_ \bot  ,m,s} \right\rangle }  + \nonumber \\
  &+& \sum\limits_{k_ \bot  ,m,s} {\psi _{k_ \bot  ,m,s}^{\left( i \right)} P\left( {k_ \bot  } \right)|\left. {k_ \bot  ,m + 2s, - s} \right\rangle }
\end{eqnarray}
where a suitable index relabelling has been performed in the second term of the RHS.

\section{The scattered electromagnetic field} \label{APP-field}
From the relation
\begin{equation}
\psi _{k_ \bot  ,m,s}^{\left( i \right)}  = \left\langle {{k_ \bot  ,m,s}}  \mathrel{\left | {\vphantom {{k_ \bot  ,m,s} {\Psi ^{\left( i \right)} }}} \right. \kern-\nulldelimiterspace} {{\Psi ^{\left( i \right)} }} \right\rangle,
\end{equation}
by exploiting Eq.(\ref{vortex}) and the identification of the $s=+1$ and $s=-1$ spin eigenvalues with the left hand and right hand circular polarizations $L$ and $R$, respectively, we get
\begin{eqnarray} \label{psi}
\begin{pmatrix}
   {\psi _{k_ \bot  ,m, + 1}^{\left( i \right)} }  \\
   {\psi _{k_ \bot  ,m, - 1}^{\left( i \right)} }  \\
\end{pmatrix} &=&
 \sqrt {\frac{{k_ \bot  }}{{2\pi }}} \int\limits_0^{ + \infty } {dr_ \bot  \;} r_ \bot  \;J_m \left( {k_ \bot  r_ \bot  } \right)\int\limits_0^{2\pi } {d\varphi } e^{ - im\varphi } \times \nonumber \\
&\times& \begin{pmatrix}
   {E_L^{\left( i \right)} \left( {r_ \bot  ,\varphi } \right)}  \\
   {E_R^{\left( i \right)} \left( {r_ \bot  ,\varphi } \right)}  \\
\end{pmatrix}.
\end{eqnarray}
The circularly polarized components of the field ${\bf{E}}_ \bot  \left( {{\bf{r}}_ \bot  } \right)$ are straightforwardly obtained by projecting the state $\left| \Psi  \right\rangle$ of Eq.(\ref{LS-Sol}) onto the basis $\left| {{\bf{r}}_ \bot  ,s} \right\rangle  = \left| {{\bf{r}}_ \bot  } \right\rangle  \otimes \left| s \right\rangle$, i.e.
\begin{eqnarray} \label{circ}
E_L \left( {{\bf{r}}_ \bot  } \right) &=& \left\langle {{{\bf{r}}_ \bot  , + 1}}  \mathrel{\left | {\vphantom {{{\bf{r}}_ \bot  , + 1} \Psi }} \right. \kern-\nulldelimiterspace} {\Psi } \right\rangle, \nonumber \\
E_R \left( {{\bf{r}}_ \bot  } \right) &=& \left\langle {{{\bf{r}}_ \bot  , - 1}} \mathrel{\left | {\vphantom {{{\bf{r}}_ \bot  , - 1} \Psi }} \right. \kern-\nulldelimiterspace}  {\Psi } \right\rangle.
\end{eqnarray}
The longitudinal component $E_z$ is discontinuous across the graphene plane (due to the surface charge oscillation hosted by graphene) and its value on the right-side at $z=0^+$ can be evaluated from the second of Eqs.(\ref{scat}). However, it is more convenient resorting to the full vectorial angular spectrum representation of the forward propagating field since it also provides the related magnetic field $\bf H$ (which we need for evaluating the Poynting vector). Considering the 2D Fourier transform of the transverse part of the field at $z=0$
\begin{equation} \label{2dfour}
{\bf{\tilde E}}_ \bot  \left( {{\bf{k}}_ \bot  } \right) = \frac{1}{{\left( {2\pi } \right)^2 }}\int {d^2 {\bf{r}}_ \bot  } e^{ - i{\bf{k}}_ \bot   \cdot {\bf{r}}_ \bot  } {\bf{E}}_ \bot  \left( {{\bf{r}}_ \bot  } \right),
\end{equation}
the forward propagating electric and magnetic fields are
\begin{widetext}
\begin{eqnarray} \label{angulspec}
\begin{pmatrix}
   {E_x }  \\
   {E_y }  \\
   {E_z }  \\
\end{pmatrix} &=& \int {d^2 {\bf{k}}_ \bot  } e^{i{\bf{k}}_ \bot   \cdot {\bf{r}}_ \bot  } \frac{{e^{ik_z z} }}{{k_z }}
 \begin{pmatrix}
   {k_z } & 0  \\
   0 & {k_z }  \\
   { - k_x } & { - k_y }  \\
\end{pmatrix}
\begin{pmatrix}
   {\tilde E_x }  \\
   {\tilde E_y }  \\
\end{pmatrix}, \nonumber \\
\begin{pmatrix}
   {H_x }  \\
   {H_y }  \\
   {H_z }  \\
\end{pmatrix} &=& \frac{1}{{\omega \mu _0 }}\int {d^2 {\bf{k}}_ \bot  } e^{i{\bf{k}}_ \bot   \cdot {\bf{r}}_ \bot  } \frac{{e^{ik_z z} }}{{k_z }} \begin{pmatrix}
   { - k_x k_y } & { - \left( {k_y^2  + k_z^2 } \right)}  \\
   {\left( {k_x^2  + k_z^2 } \right)} & {k_x k_y }  \\
   { - k_y k_z } & {k_x k_z }  \\
\end{pmatrix} \begin{pmatrix}
   {\tilde E_x }  \\
   {\tilde E_y }  \\
\end{pmatrix},
\end{eqnarray}
\end{widetext}
where $k_z = \sqrt{k_0^2-k_\bot^2}$. Inserting Eqs.(\ref{circ}) into Eq.(\ref{2dfour}), after using circularly polarized components ${\bf{\tilde E}}_ \bot   = \tilde E_L {\bf{e}}_L  + \tilde E_R {\bf{e}}_R$, introducing polar coordinates in both direct and reciprocal space by setting ${\bf{r}}_ \bot   = r_ \bot  \left( {\cos \varphi {\bf{e}}_x  + \sin \varphi {\bf{e}}_y } \right)$, ${\bf{k}}_ \bot   = k_ \bot  \left( {\cos \theta {\bf{e}}_x  + \sin \theta {\bf{e}}_y } \right)$ and resorting to the Anger-Jacobi formula
\begin{equation}
e^{i\zeta \cos \Phi }  = \sum\limits_{m =  - \infty }^{ + \infty } {i^m } J_m \left( \zeta  \right)e^{im\Phi },
\end{equation}
we get
\begin{equation} \label{LR}
\begin{pmatrix}
   {\tilde E_L }  \\
   {\tilde E_R }  \\
\end{pmatrix} = \sum\limits_{m =  - \infty }^\infty  {\frac{{e^{im\theta } }}{{i^m \sqrt {8\pi ^3 k_ \bot  } }}} \begin{pmatrix}
   {Q\psi _{k_ \bot  ,m, + 1}^{\left( i \right)}  + P\psi _{k_ \bot  ,m + 2, - 1}^{\left( i \right)} }  \\
   {P\psi _{k_ \bot  ,m - 2, + 1}^{\left( i \right)}  + Q\psi _{k_ \bot  ,m, - 1}^{\left( i \right)} }  \\
\end{pmatrix}.
\end{equation}
Using circularly polarized components for the transverse parts of the fields in Eqs.(\ref{angulspec}) at $z=0$ and using Eq.(\ref{LR}), after some tedious but straightforward algebra, we obtain
\begin{widetext}
\begin{eqnarray} \label{FIELDS}
 \begin{pmatrix}
   {E_L }  \\
   {E_R }  \\
   {E_z }  \\
\end{pmatrix} &=& \int\limits_0^{ + \infty } {dk_ \bot  } \sum\limits_{m =  - \infty }^\infty  {\sqrt {\frac{{k_ \bot  }}{{2\pi }}} \begin{pmatrix}
   {J_m \left( {k_ \bot  r_ \bot  } \right)e^{im\varphi } Q} & {J_{m - 2} \left( {k_ \bot  r_ \bot  } \right)e^{i\left( {m - 2} \right)\varphi } P}  \\
   {J_{m + 2} \left( {k_ \bot  r_ \bot  } \right)e^{i\left( {m + 2} \right)\varphi } P} & {J_m \left( {k_ \bot  r_ \bot  } \right)e^{im\varphi } Q}  \\
   {J_{m + 1} \left( {k_ \bot  r_ \bot  } \right)e^{i\left( {m + 1} \right)\varphi } \frac{{i  }}{{k_z }}W_ - } & { - J_{m - 1} \left( {k_ \bot  r_ \bot  } \right)e^{i\left( {m - 1} \right)\varphi } \frac{{i  }}{{k_z }} W_ -}  \\
\end{pmatrix}\begin{pmatrix}
   {\psi _{k_ \bot  ,m, + 1}^{\left( i \right)} }  \\
   {\psi _{k_ \bot  ,m, - 1}^{\left( i \right)} }  \\
\end{pmatrix}}, \nonumber  \\
 \begin{pmatrix}
   {H_L }  \\
   {H_R }  \\
   {H_z }  \\
\end{pmatrix} &=& \frac{1}{{\omega \mu _0 }}\int\limits_0^{ + \infty } {dk_ \bot  } \sum\limits_{m =  - \infty }^\infty  {\sqrt {\frac{{k_ \bot  }}{{2\pi }}} } \begin{pmatrix}
   { - J_m \left( {k_ \bot  r_ \bot  } \right)e^{im\varphi } \tilde Q} & { - J_{m - 2} \left( {k_ \bot  r_ \bot  } \right)e^{i\left( {m - 2} \right)\varphi } \tilde P}  \\
   {J_{m + 2} \left( {k_ \bot  r_ \bot  } \right)e^{i\left( {m + 2} \right)\varphi } \tilde P} & {J_m \left( {k_ \bot  r_ \bot  } \right)e^{im\varphi } \tilde Q}  \\
   {J_{m + 1} \left( {k_ \bot  r_ \bot  } \right)e^{i\left( {m + 1} \right)\varphi } W_ +  } & {J_{m - 1} \left( {k_ \bot  r_ \bot  } \right)e^{i\left( {m - 1} \right)\varphi } W_ +  }  \\
\end{pmatrix}\begin{pmatrix}
   {\psi _{k_ \bot  ,m, + 1}^{\left( i \right)} }  \\
   {\psi _{k_ \bot  ,m, - 1}^{\left( i \right)} }  \\
\end{pmatrix},
\end{eqnarray}
\end{widetext}
where
\begin{eqnarray}
 W_ \pm   &=&  - \frac{{k_ \bot  }}{{\sqrt 2 }}\left[ {Q \pm P} \right], \nonumber \\
 \tilde Q &=& \frac{i}{{2k_z }}\left[ {\left( {k_z^2  + k_0^2 } \right)Q - k_ \bot ^2 P} \right], \nonumber \\
 \tilde P &=& \frac{i}{{2k_z }}\left[ { - k_ \bot ^2 Q + \left( {k_z^2  + k_0^2 } \right)P} \right].
\end{eqnarray}
Equations (\ref{FIELDS}) together with Eq.(\ref{psi}) allows to predict the full electromagnetic field on the right side of the graphene sheet (i.e. at $z=0^+$) once the transverse part of the incoming field is known.


\begin{thebibliography}{10}
\newcommand{\enquote}[1]{``#1''}

\bibitem{Allen1992} L. Allen, M. W. Beijersbergen, R. Spreeuw, and J. Woerdman, {\it Phys. Rev. A} {\bf 45}, 8185--8189 (1992).

\bibitem{Mair2001} A. Mair, A. Vaziri, G. Weihs, and A. Zeilinger, {\it Nature} {\bf 412}, 313--316 (2001).

\bibitem{Nagali2010} E. Nagali, L. Sansoni, L. Marrucci, E. Santamato, and F. Sciarrino, {\it Phys. Rev. A} {\bf 81}, 052317 (2010).

\bibitem{Molina2007} G. Molina-Terriza, J. P. Torres, and L. Torner, {\it Nat. Phys.} {\bf 3}, 305--310 (2007).

\bibitem{Cardano2015} F. Cardano, F. Massa, H. Qassim, E. Karimi, S. Slussarenko, D. Paparo, C. de Lisio, F. Sciarrino, E. Santamato, R. W. Boyd, and L. Marrucci, {\it Sci. Adv.} {\bf 1}, e1500087 (2015).

\bibitem{Bechmann2000} H. Bechmann-Pasquinucci, and W. Tittel, {\it Phys. Rev. A} {\bf 61}, 062308 (2000).

\bibitem{Cerf2002} N. J. Cerf, M. Bourennane, A. Karlsson, and N. Gisin, {\it Phys. Rev. Lett.} {\bf 88}, 127902 (2002).

\bibitem{Mirho2015} M. Mirhosseini, O. S. Magana-Loaiza, M. N. O'Sullivan, B. Rodenburg, M. Malik, M. P. Lavery, M. J. Padgett, D. J. Gauthier, and R. W. Boyd, {\it New J. Phys.} {\bf 17}, 033033 (2015).

\bibitem{Paterson2005} C. Paterson, {\it Phys. Rev. Lett.} {\bf 94}, 153901 (2005).

\bibitem{MalikOE2012} M. Malik, M. O'Sullivan, B. Rodenburg, M. Mirhosseini, J. Leach, M. P. Lavery, M. J. Padgett, and R. W. Boyd, {\it Opt. Express} {\bf 20}, 13195--13200 (2012).

\bibitem{Willner2015} A. E. Willner, H. Huang, Y. Yan, Y. Ren, N. Ahmed, G. Xie, C. Bao, L. Li, Y. Cao, Z. Zhao, J. Wang, M. P. J. Lavery, M. Tur, S. Ramachandran, A. F. Molisch, N. Ashrafi, and S. Ashrafi, {\it Adv. Opt. Photon.} {\bf 7}, 66--106 (2015).

\bibitem{FarasSR2015} O. J. Faras, V. D'Ambrosio, C. Taballione, F. Bisesto, S. Slussarenko, L. Aolita, L. Marrucci, S. P. Walborn, and F. Sciarrino, {\it Sci. Rep.} {\bf 5}, 8424 (2015).

\bibitem{Goyal2016} S. K. Goyal, A. H. Ibrahim, F. S. Roux, T. Konrad, and A. Forbes, {\it J. Opt.} {\bf 18}, 064002 (2016).

\bibitem{TamburiniNJP2012} F. Tamburini, E. Mari, A. Sponselli, B. Thide, A. Bianchini, and F. Romanato, {\it New J. Phys.} {\bf 14}, 033001 (2012).

\bibitem{WangNP2012} J. Wang, J.-Y. Yang, I. M. Fazal, N. Ahmed, Y. Yan, H. Huang, Y. Ren, Y. Yue, S. Dolinar, M. Tur, and A. E. Willner, {\it Nat. Photonics} {\bf 6}, 488--496 (2012).

\bibitem{Krenn2014} M. Krenn, R. Fickler, M. Fink, J. Handsteiner, M. Malik, T. Scheidl, R. Ursin, and A. Zeilinger, {\it New J. Phys.} {\bf 16}, 113028 (2014).

\bibitem{KrennPNAS2016} M. Krenn, J. Handsteiner, M. Fink, R. Fickler, R. Ursin, M. Malik, and A. Zeilinger, {\it Proc. Natl. Acad. Sci. U.S.A.} {\bf 113}, 13648--13653 (2016).

\bibitem{Pfeifer2007} R. N. C. Pfeifer, T. A. Nieminen, N. R. Heckenberg, and H. Rubinsztein-Dunlop, {\it Rev. Mod. Phys.} {\bf 79}, 1197--1216 (2007).

\bibitem{Chu1998} S. Chu, {\it Rev. Mod. Phys.} {\bf 70}, 685--706 (1998).

\bibitem{Phillips1998} W. D. Phillips, {\it Rev. Mod. Phys.} {\bf 70}, 721--741 (1998).

\bibitem{Ashkin2000} A. Ashkin, {\it IEEE J. Sel. Top. Quantum Electron.} {\bf 6}, 841--856 (2000).

\bibitem{Grier2003} D. G. Grier, {\it Nature} {\bf 424}, 810--816 (2003).

\bibitem{Dienero2008} M. Dienerowitz, M. Mazilu, and K. Dholakia, {\it J. Nanophoton.} {\bf 2}, 021875 (2008).

\bibitem{Bowman2013} R. W. Bowman, and M. J. Padgett, {\it Rep. Prog. Phys.} {\bf 76}, 026401 (2013).

\bibitem{MarrucciJOPT2011} L. Marrucci, E. Karimi, S. Slussarenko, B. Piccirillo, E. Santamato, E. Nagali, and F. Sciarrino, {\it J. Opt.} {\bf 13}, 064001 (2011).

\bibitem{Zayats2015} K. Y. Bliokh, F. J. Rodriguez-Fortuno, F. Nori, and A. V. Zayats, {\it Nat. Photon.} {\bf 9}, 796--808 (2015).

\bibitem{Mathur1991} H. Mathur, {\it Phys. Rev. Lett.} {\bf 67}, 3325--3327 (1991).

\bibitem{Xiao2010} D. Xiao, M.-C. Chang, and Q. Niu, {\it Rev. Mod. Phys.} {\bf 82}, 1959--2007 (2010).

\bibitem{Zhao2007} Y. Zhao, J. S. Edgar, G. D. M. Jeffries, D. McGloin, and D. T. Chiu, {\it Phys. Rev. Lett.} {\bf 99}, 073901 (2007).

\bibitem{Kavokin2005} A. Kavokin, G. Malpuech, and M. Glazov, {\it Phys. Rev. Lett.} {\bf 95}, 136601 (2005).

\bibitem{Hosten2008} O. Hosten, and P. Kwiat, {\it Science} {\bf 319}, 787-790 (2008).

\bibitem{Ling2017} X. Ling, X. Zhou, K. Huang, Y. Liu, C. W. Qiu, H. Luo, and S. Wen, {\it Rep. Prog. Phys.} {\bf 80}, 066401 (2017).

\bibitem{Bliokh2006} K. Y. Bliokh, and Y. P. Bliokh, {\it Phys. Rev. Lett.} {\bf 96}, 073903 (2006).

\bibitem{Bliokh2013} K. Y. Bliokh, and A. Aiello, {\it J. Opt.} {\bf 15} 014001 (2013).

\bibitem{Aiello2012} A. Aiello, {\it New Jour. of Phys.} {\bf 14}, 013058 (2012).

\bibitem{Ciattoni2003} A. Ciattoni, G. Cincotti, and C. Palma, {\it J. Opt. Soc. Am A} {\bf 20}, 163-171 (2003).

\bibitem{Brasselet2009} E. Brasselet, Y. Izdebskaya, V. Shvedov, A. S. Desyatnikov, W. Krolikowski, and Y. S. Kivshar, {\it Opt. Lett.} {\bf 34}, 1021-1023 (2009).

\bibitem{Yavorsky2012} M. Yavorsky, and E. Brasselet, {\it Opt. Lett.} {\it 37}, 3810-3812 (2012).

\bibitem{Ciattoni2017} A. Ciattoni, A. Marini, and C. Rizza, {\it Phys. Rev. Lett.} {\bf 118}, 104301 (2017).    

\bibitem{OConnor2014} D. O'Connor, P. Ginzburg, F.J. Rodriguez-Fortuno, G. A. Wurtz, and A. V. Zayats, {\it Nat. Commun.} {\bf 5}, 5327 (2014).

\bibitem{Pan2016} D. Pan, H. Wei, L. Gao, and H. Xu, {\it Phys. Rev. Lett.} {\bf 117}, 166803 (2016).

\bibitem{Rodrig2013} F. J. Rodriguez-Fortuno, G, Marino, P. Ginzburg, D. O'Connor, A. Martinez, G. A. Wurtz, and A. V. Zayats, {\it Science} {\bf 340}, 328-–330 (2013).

\bibitem{Geim2007} A. K. Geim, and K. S. Novoselov, {\it Nat. Mater.} {\bf 6}, 183--191 (2007).

\bibitem{Bonaccorso2010} F. Bonaccorso, Z. Sun, T. Hasan, and A. C. Ferrari, {\it Nat. Photon.} {\bf 4}, 611--622 (2010).

\bibitem{JavierACSPhot} F. J. Garc\'{\i}a de Abajo, {\it ACS Photonics} {\bf 1} 135--152 (2014).

\bibitem{Koppens2011} F. H. L. Koppens, D. E. Chang, and F.J. Garc\'{\i}a de Abajo, {\it Nano Lett.} {\bf 11} 3370--3377 (2011).

\bibitem{Li2014} Y. Li, H. Yan, D. B. Farmer, X. Meng, W. Zhu, R. M. Osgood, T. F. Heinz, and P. Avouri, {\it Nano Lett.} {\bf 14}, 1573--1577 (2014).

\bibitem{Rodrigo2015} D. Rodrigo, O. Limaj, D. Janner, D. Etezadi, F. J. Garc\'{\i}a de Abajo, V. Pruneri, and H. Altug, {\it Science} {\bf 349}, 165--168 (2015).

\bibitem{Marini2015} A. Marini, I. Silveiro, and F. J. Garc\'{\i}a de Abajo, {\it ACS Photonics} {\bf 2} 876--882 (2015).

\bibitem{Mikhailov2011} S. A. Mikhailov, {\it Phys. Rev. B} {\bf 84}, 045432 (2011).

\bibitem{Cox2017} J. D. Cox, A. Marini, and F. J. Garc\'{\i}a de Abajo, {\it Nat. Commun.} {\bf 8} 14380 (2017).

\bibitem{Sukosin2012} S. Thongrattanasiri, F. H. L. Koppens, and F. J. Garc\'{\i}a de Abajo, {\it Phys. Rev. Lett.} {\bf 108} 047401 (2012).

\bibitem{Berardi2012} B. Sensale-Rodriguez, R. Yan, M. M. Kelly, T. Fang, K. Tahy, W. S. Hwang, D. Jena, L. Liu, and H. G. Xing, {\it Nat. Commun.} {\bf 3}, 780 (2012).

\bibitem{Berardi2013} B. Sensale-Rodriguez, R. Yan, S. Rafique, M. Zhu, W. Li, X. Liang, D. Gundlach, V. Protasenko, M. M. Kelly, D. Jena, L. Liu, and H. G. Xing, {\it Nano Lett.} {\bf 12}, 4518--4522 (2013).

\bibitem{Schwinger}  J. Schwinger, L. L. Deraad Jr., K. A. Milton, W. Tsai, and J. Norton, {\it Classical Electrodynamics}, CRC Press; First Edition (11 Sept. 1998).

\bibitem{App_Mandle} L. Mandel, and Emil Wolf, {\it Optical Coherence and Quantum Optics} Cambridge University Press, 1995.

\bibitem{CastroNeto2009} A. H. Castro Neto, F. Guinea, N. M. R. Peres, K. S. Novoselov, and A. K. Geim, {\it Rev. Mod. Phys.} {\bf 81}, 109--162 (2009).

\bibitem{Chen2011} C.-F. Chen, C.-H. Park, B. W. Boudouris, J. Horng, B. Geng, C. Girit, A. Zettl,  M. F. Crommie, R. A. Segalman, S. G. Louie, and F. Wang, {\it Nature} {\bf 471}, 617--620 (2011).

\bibitem{Liu2011} H. Liu, Y. Liu, and D. Zhua, {\it J. Mater. Chem.} {\bf 21}, 3335--3345 (2011).

\bibitem{Karimi2014} E. Karimi, S. A. Schulz, I. De Leon, H. Qassim, J. Upham, and R. W Boyd, {\it Light Sci. Appl.} {\bf 3}, e167 (2014).

\end{thebibliography}
\end{document}